\documentclass[graybox]{svmult}

\usepackage{mathptmx}       
\usepackage{helvet}         
\usepackage{courier}        
\usepackage{makeidx}         
\usepackage{graphicx}        
\usepackage{multicol}        
\usepackage[bottom]{footmisc}

\usepackage{color}
\definecolor{keywordcolor}{rgb}{0,0,0} 
\providecommand{\keyword}[1]{\index{#1}{\color{keywordcolor}#1}}

\makeindex             

\begin{document}

\title*{Cavity Induced Interfacing of Atoms and Light}
\author{Axel Kuhn
}
\institute{Axel Kuhn \at University of Oxford, Clarendon Laboratory, Parks Road, Oxford, UX1 3PU, UK \email{axel.kuhn@physics.ox.ac.uk}
}
%
%
\maketitle

\abstract{This chapter introduces cavity-based light-matter \keyword{quantum interfaces}, with a single atom or ion in strong coupling to a high-finesse optical \keyword{cavity}. We discuss the deterministic generation of indistinguishable \keyword{single photons} from these systems; the \keyword{atom-photon entanglement} intractably linked to this process; and the information encoding using spatio-temporal modes within these photons. Furthermore, we show how to establish a time-reversal of the aforementioned emission process to use a coupled atom-cavity system as a quantum memory. Along the line, we also discuss the performance and characterisation of cavity photons in elementary linear-optics arrangements with single beam splitters for quantum-homodyne measurements.}

\section{Introduction}
The interfacing of discrete matter states and photons, the storage and retrieval of single photons, and the \keyword{entanglement} and mapping of quantum states between distant entities are key elements of quantum networks for distributed quantum information processing \cite{DiVincenzo98}.  Ideally, such systems are composed of individual nodes acting as quantum gates or memories, with optical interlinks that allow for the entanglement or teleportation of their quantum states, or for optical quantum information processing using linear optics acting on the light traveling between the nodes \cite{Knill01,Reiserer}. With individual photons carrying the information, substantial efforts have been made that focus on their production and characterization. Applications that rely on single photons and on their indistinguishability include quantum cryptography, optical quantum computing, light-matter entanglement, and atom-photon state mapping, which all have successfully been demonstrated. For these purposes, sources of single photons that are based on single isolated quantum systems like a single atom or ion are ideal, given their capability of emitting streams of indistinguishable photons on demand. This approach is inherently simple and robust because a single quantum system can only emit one single photon in a de-excitation process.  With all atoms or ions of the same isotope being identical, different photon sources based on one-and-the-same species are able of producing indistinguishable photons without further measures, provided the same transitions are used and the electromagnetic environment is identical for all atoms. This makes them ideal candidates for the implementation of large-scale quantum computing networks. 

In the present chapter, we primarily discuss the quantum-state control of an atom strongly coupled to an optical cavity, with particular focus on the deterministic generation of single photons in arbitrary spatio-temporal modes.  Important fundamental properties of on-demand single photon sources are analyzed, including single photon purity and indistinguishability. In this context, understanding and controlling the fundamental processes that govern the interaction of atoms with optical cavities is important for the development of improved single-photon emitters.  These interactions are examined in the context of cavity-quantum electrodynamic (\keyword{cavity-QED}) effects. We show how to apply these to channel the photon emission into a single mode of the radiation field, with the vacuum field inside the cavity stimulating the process. Furthermore we elucidate how to determine and control the coherence properties of these photons in the time domain and use that degree of control for information encoding, and for information or photon-storage in single atoms by means of a time-reversal of the photon emission processes.

\section{Cavities for interfacing light and matter}
In this section, we closely follow, summarize and extend our recently published review articles \cite{Kuhn10,Solomon13} in order to introduce the concepts, characteristic properties, and major implementations of state-of-the-art single-photon sources based on single atoms or ions in cavities. These have all the potential to meet the requirements of optical quantum computing and quantum networking schemes, namely deterministic single-photon emission with unit efficiency, directed emission into a single spatial mode of the radiation field, indistinguishable photons with immaculate temporal and spatial coherence, and reversible quantum state mapping and entanglement between atoms and photons.

Starting from the elementary principles of cavity quantum electrodynamics, we discuss the coupling of a single quantum system to the quantised radiation field within optical resonators. Then we show how to exploit these effects to generate single photons on demand in the strong coupling regime and the bad cavity limit, using either an adiabatic driving technique or a sudden excitation of the emitter. To conclude, we discuss a couple of prominent experimental achievements and examine the different approaches for obtaining single photons from cavities using either atoms or ions as photon emitters.

\subsection{Atom-photon interaction in resonators}
\label{sec:atom-photon interaction}
Any single quantum system that shows discrete energy levels, like an individual atom or ion, can be coupled to the quantised modes of the radiation field in a cavity. Here we  introduce the relevant features of \keyword{cavity-QED} and the \keyword{Jaynes-Cummings model} \cite{Jaynes63,Shore93}, and then extend these to three-level atoms with two dipole transitions driven by two radiation fields. One of the fields is from a laser, the other is the cavity field  coupled to the atom.  We furthermore explain how the behaviour of a coupled-atom system depends on the most relevant cavity parameters, such as the cavity's mode volume and its finesse.\\

\noindent{\bf Field quantisation:}
We consider  a Fabry-Perot cavity with mirror separation $l$ and reflectivity $\mathcal{R}$. The cavity has a free spectral range ${\Delta\omega_\mathrm{FSR}=2\pi \times c /(2l)}$, and its finesse is defined as ${\mathcal{F}={\pi\sqrt{\mathcal{R}}}/{(1-\mathcal{R})}}$. In the vicinity of a resonance, the transmission profile is Lorentzian with a linewidth (FWHM) of ${2\kappa=\Delta\omega_\mathrm{FSR}/\mathcal{F}}$, which is twice the decay rate, $\kappa$, of the cavity field. Curved mirrors are normally used to restrict the cavity eigenmodes to geometrically stable Laguerre-Gaussian or Hermite-Gaussian modes. In most cases, just one of these modes is of interest, characterised by its mode function $\psi_\mathrm{cav}(\mathbf{r})$ and its resonance frequency $\omega_\mathrm{cav}$. The state vector can therefore be expressed as a superposition of photon-number states, $|n\rangle$, and for $n$ photons in the mode the energy is ${\hbar\omega_\mathrm{cav}(n+\frac{1}{2})}$. The equidistant energy spacing imposes an analogous treatment of the cavity as a harmonic oscillator. Creation and annihilation operators for a photon, $\hat{a}^\dag$ and $\hat{a}$, are then used to express the Hamiltonian of the cavity,
\begin{equation}
    H_\mathrm{cav}=\hbar\omega_\mathrm{cav}\left(\hat{a}^\dag \hat{a}+\frac{1}{2}\right).
\end{equation}
We emphasize that this does not take account of any losses, whereas in a real cavity, all photon number states decay until thermal equilibrium with the environment is reached. In the optical domain, the latter corresponds to the vacuum state, $|0\rangle$, with no photons remaining in the cavity.\\

\noindent{\bf Two-level atom:}
We now analyse how the cavity field interacts with a two-level atom with ground state $|g\rangle$ and excited state $|x\rangle$ of energies $\hbar\omega_\mathrm{g}$ and $\hbar\omega_\mathrm{x}$, respectively, and transition dipole moment $\mu_\mathrm{xg}$. The Hamiltonian of the atom reads 
\begin{equation}
  H_\mathrm{A}=\hbar\omega_\mathrm{g}|g\rangle\langle g| + \hbar\omega_\mathrm{x}|x\rangle\langle x|.
\end{equation}
The coupling to the field mode of the cavity is expressed by the atom-cavity coupling constant,
\begin{equation}
    g(\mathbf{r})=g_\mathrm{0} \ \psi_\mathrm{cav}(\mathbf{r}), 
    \ \mathrm{with}\ 
    g_\mathrm{0}=\sqrt{(\mu_\mathrm{xg}^2\omega_\mathrm{cav})/(2\hbar\epsilon_\mathrm{0}V)},
\end{equation}
where $V$ is the mode volume of the cavity.  As the atom is barely moving during the interaction, we can safely disregard its external degrees of freedom. Furthermore we assume maximum coupling, i.e. $ \psi_\mathrm{cav}(\mathbf{r}_\mathrm{atom})=1$, so that one obtains $g(\mathbf{r})=g_\mathrm{0}$. In a closed system, any change of the atomic state would go hand-in-hand with a corresponding change of the photon number, $n$.  Hence the interaction Hamiltonian of the atom-cavity system reads 
\begin{equation}
    H_\mathrm{int}=-\hbar g_\mathrm{0}
    \left[|x\rangle\langle g| \hat{a} + \hat{a}^\dag |g\rangle\langle x|\right].
\end{equation}
For a given excitation number $n$, the cavity only couples $|g,n\rangle$ and $|x,n-1\rangle$. If the cavity mode is resonant with the atomic transition, i.e. if $\omega_{cav}=\omega_x-\omega_g$, the population oscillates with the Rabi frequency $\Omega_\mathrm{cav}=2g_\mathrm{0}\sqrt{n}$ between these states. 

\begin{figure}[htb]
	\centering\includegraphics[width=4in]{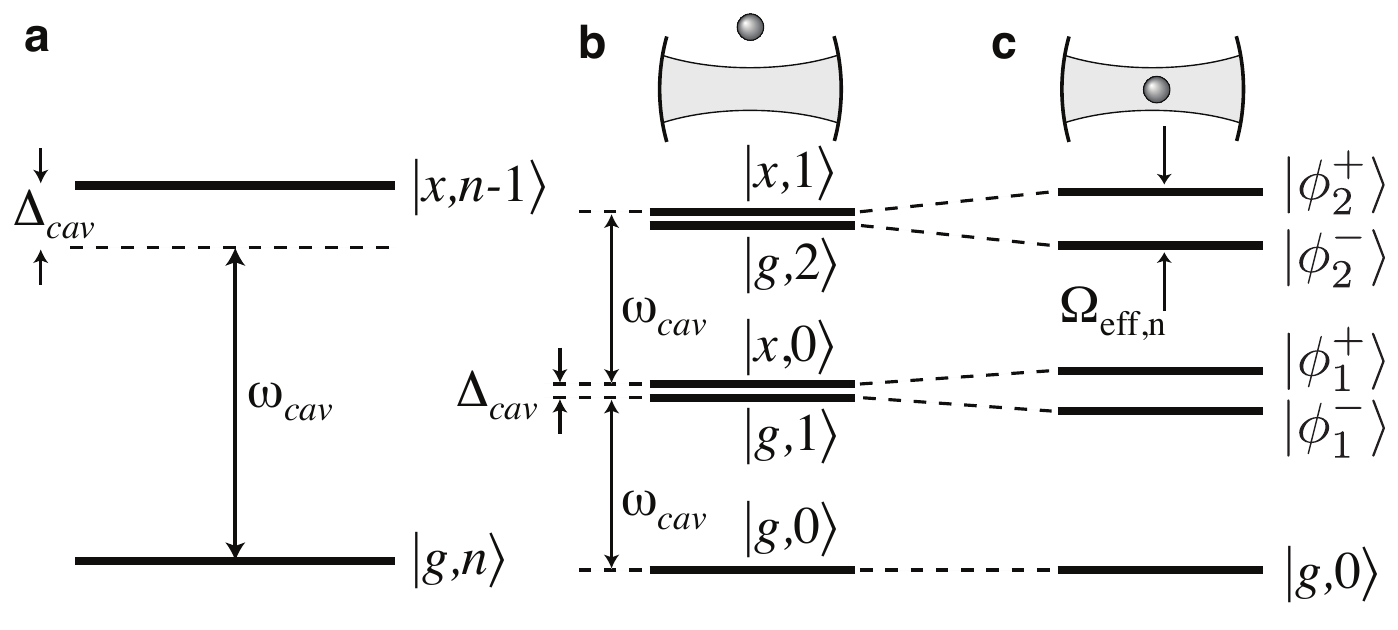}
	\caption{Atom-cavity coupling (from \cite{Kuhn10}): \textbf{(a)} A two-level atom with ground state $|g\rangle$   and excited state $|x\rangle$ coupled to a cavity containing $n$   photons. In the dressed-level scheme of the combined atom-cavity   system with the atom outside \textbf{(b)} or inside \textbf{(c)} the   cavity, the state doublets are either split by $\Delta_\mathrm{cav}$ or by   the effective Rabi frequency, $\Omega_\mathrm{eff,n}$,  respectively.}
	\label{fig:twolevelscheme}
\end{figure}

The eigenfrequencies of the total Hamiltonian, $H=H_\mathrm{cav}+H_\mathrm{A}+H_\mathrm{int}$, can be found easily. In the rotating wave approximation, they read
\begin{equation}
    \omega^\pm_\mathrm{n}=\omega_\mathrm{cav}\left(n+\frac{1}{2}\right) + 
    \frac{1}{2}\left(\Delta_\mathrm{cav}\pm\sqrt{4 n g_\mathrm{0}^2 + 
    \Delta_\mathrm{cav}^2}\right),
    \label{eq:opmcav}
\end{equation}
where $\Delta_\mathrm{cav}=\omega_\mathrm{x}-\omega_\mathrm{g}-\omega_\mathrm{cav}$ is the detuning between atom and cavity. 
Fig.\,\ref{fig:twolevelscheme} illustrates this level splitting. Within each $n$ manifold, the two eigenstates are split by  $\Omega_\mathrm{eff,n}=\sqrt{4 n   g_\mathrm{0}^2+\Delta_\mathrm{cav}^2}$, which is the effective Rabi frequency at which the population oscillates between states $|g,n\rangle$ and $|x,n-1\rangle$. This means that the cavity field stimulates the emission of an excited atom into the cavity, thus de-exciting the atom and increasing the photon number by one. Subsequently, the atom is re-excited by absorbing a photon from the cavity field, and so forth. In particular, an excited atom and a cavity containing no photon are sufficient to start the oscillation between $|x,0\rangle$ and $|g,1\rangle$ at frequency $\sqrt{4 g_\mathrm{0}^2+\Delta_\mathrm{cav}^2}$. This phenomenon is known as vacuum Rabi oscillation. On resonance, i.e. for $\Delta_\mathrm{cav}=0$, the oscillation frequency is $2g_\mathrm{0}$, also known as vacuum Rabi frequency.

To summarise, the atom-cavity interaction splits the photon number states into doublets of non-degenerate dressed states, which are named after Jaynes and Cummings \cite{Jaynes63,Shore93}. Only the ground state $|g,0\rangle$ is not coupled to other states and is  not subject to any energy shift or splitting.\\

\begin{figure}
	\centering\includegraphics[width=4in]{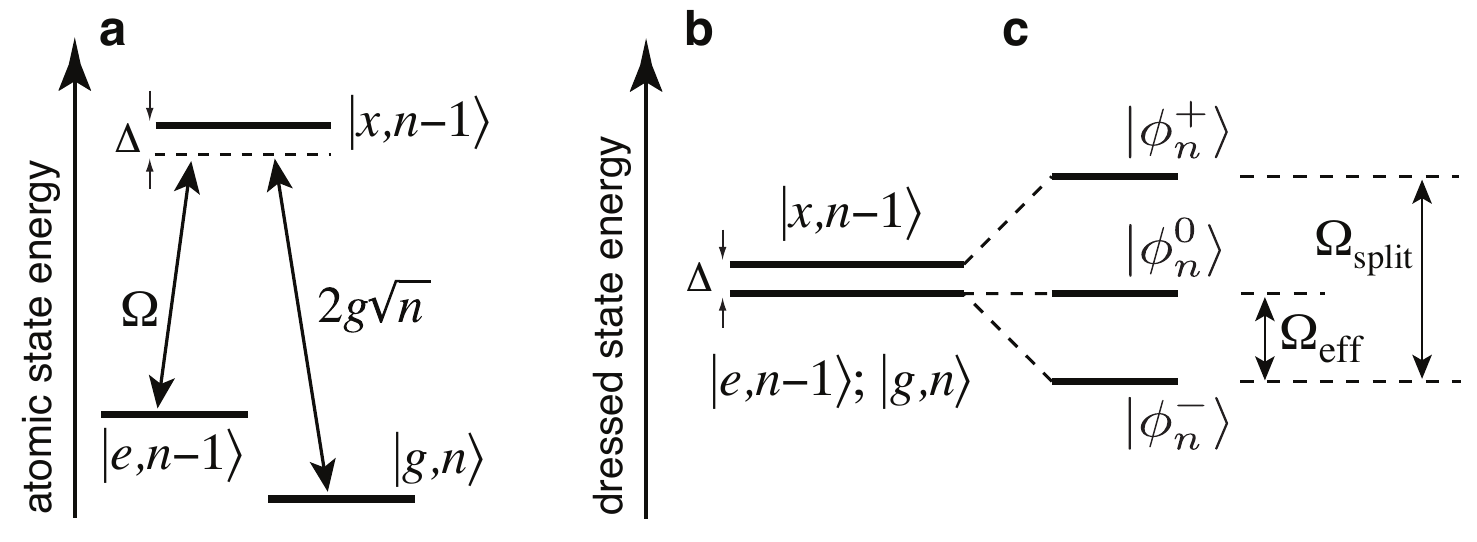}
	\caption{Three-level atom in cavity coupling (from \cite{Kuhn10}): \textbf{(a)} A three-level atom driven by a classical laser   field of Rabi frequency $\Omega$, coupled to a cavity containing $n$   photons.  \textbf{(b)} Dressed-level scheme of the combined system   without coupling, and \textbf{(c)} for an atom interacting with   laser and cavity. The triplet is split by   $\Omega_\mathrm{split}=\sqrt{4 n g_\mathrm{0}^2+\Omega^2+\Delta^2}$. In the   limit of a large detuning $\Delta$, the Raman transition   $|e,n-1\rangle\leftrightarrow|g,n\rangle$ is driven at the effective   Rabi frequency   $\Omega_\mathrm{eff}=\frac{1}{2}\left(\Omega_\mathrm{split}-|\Delta|\right)\approx (4 n g_\mathrm{0}^2+\Omega^2)/|4\Delta|$.}
	\label{fig:threelevelscheme}
\end{figure}

\noindent{\bf Three-level atom:}
\label{sec:three-level-atom}
We now consider an atom with a $\Lambda-$type three-level scheme providing transition frequencies $\omega_\mathrm{xe}=\omega_\mathrm{x}-\omega_\mathrm{e}$ and $\omega_\mathrm{xg}=\omega_\mathrm{x}-\omega_\mathrm{g}$ as depicted in Fig.\,\ref{fig:threelevelscheme}. The $|e\rangle \leftrightarrow |x\rangle$ transition is driven by a classical light field of frequency $\omega_\mathrm{L}$ with Rabi frequency $\Omega$, while a cavity mode with frequency $\omega_\mathrm{cav}$ couples to the $|g\rangle \leftrightarrow |x\rangle$ transition. The respective detunings are defined as $\Delta_\mathrm{L}=\omega_\mathrm{xe}-\omega_\mathrm{L}$ and $\Delta_\mathrm{cav}=\omega_\mathrm{xg}-\omega_\mathrm{cav}$. Provided the driving laser and the cavity only couple to their respective transitions, the interaction Hamiltonian
\begin{equation}
\begin{array}{rl}
H_\mathrm{int}= \hbar [&\Delta_\mathrm{L}|e\rangle\langle e| + \Delta_\mathrm{cav}|g\rangle\langle  g| - \frac{\Omega}{2} (|x\rangle\langle e| + |e\rangle\langle x| ) 
\\[1mm]
&
 - g_\mathrm{0} (|x\rangle\langle g| a + a^\dag |g\rangle\langle x|)]
\end{array}
\label{eq:Hatcav3}
\end{equation}
determines the behaviour of the system.
Given an arbitrary excitation number $n$, this Hamiltonian couples only the three states $|e,n-1\rangle$, $|x,n-1\rangle$, $|g,n\rangle$. For this triplet and a Raman-resonant interaction with $\Delta_\mathrm{L}=\Delta_\mathrm{cav}\equiv \Delta$, the eigenfrequencies of the coupled system read
\begin{eqnarray}
&&  \omega_\mathrm{n}^0=\omega_\mathrm{cav} \left(n+\frac{1}{2}\right) \quad\mbox{and}\\
&& \omega_\mathrm{n}^\pm =\omega_\mathrm{cav} \left(n+\frac{1}{2}\right) + 
  \frac{1}{2}\left(\Delta\pm\sqrt{4 n g_\mathrm{0}^2 + \Omega^2 +
      \Delta^2}\right).
\nonumber
\end{eqnarray}
The previously-discussed Jaynes-Cummings doublets are now replaced by \keyword{dressed-state} triplets,
\begin{eqnarray}
  |\phi_\mathrm{n}^0\rangle &=& \cos\Theta |e,n-1\rangle - \sin\Theta 
  |g,n\rangle\label{anull1},\\
  |\phi_\mathrm{n}^+\rangle &=& \cos\Phi\sin\Theta |e,n-1\rangle - \sin\Phi |x,n-1\rangle   + \cos\Phi\cos\Theta |g,n\rangle,\nonumber\\
  |\phi_\mathrm{n}^-\rangle &=& \sin\Phi\sin\Theta |e,n-1\rangle + \cos\Phi |x,n-1\rangle  + \sin\Phi\cos\Theta |g,n\rangle,\nonumber
\end{eqnarray}
where the \keyword{mixing angles} $\Theta$ and $\Phi$ are given by
\begin{equation}
  \tan\Theta = \frac{\Omega}{2 g_\mathrm{0} \sqrt{n}}, \quad \tan\Phi = 
  \frac{\sqrt{4 n g_\mathrm{0}^2 +
      \Omega^2}}{\sqrt{4 n g_\mathrm{0}^2+\Omega^2+\Delta^2}-\Delta}.
\end{equation}
The interaction with the light lifts the degeneracy of the  eigenstates. However, $|\phi_\mathrm{n}^0\rangle$ is neither subject to an energy shift, nor does the excited atomic state contribute to it. Therefore it is a `\keyword{dark state}' which cannot decay by spontaneous emission.

In the limit of vanishing $\Omega$, the states $|\phi^\pm_\mathrm{n}\rangle$ correspond to the Jaynes-Cummings doublet and the third eigenstate, $|\phi^0_\mathrm{n}\rangle$, coincides with $|e,n-1\rangle$. Also the eigenfrequency $\omega^0_\mathrm{n}$ is not affected by $\Omega$ or $g_\mathrm{0}$. Therefore transitions between the dark states $|\phi^0_\mathrm{n+1}\rangle$ and $|\phi^0_\mathrm{n}\rangle$ are always in resonance with the cavity.  This holds, in particular, for the transition from $|\phi^0_\mathrm{1}\rangle$ to $|\phi^0_\mathrm{0}\rangle\equiv|g,0\rangle$ as the $n=0$ state never splits.\\

\noindent{\bf Cavity-coupling regimes:}
So far, we have been considering the interaction Hamiltonian and the associated eigenvalues and dressed eigenstates that one obtains whenever a two- or three-level atom is coupled to a cavity. We have been neglecting the atomic polarisation decay rate, $\gamma$, and also the field-decay rate of the cavity, $\kappa$ (Note that we have chosen a definition where the population decay rate of the atom reads $2\gamma$, and the photon loss rate from the cavity is $2\kappa$). It is evident that both relaxation rates result in a damping of a possible vacuum-Rabi oscillation between states $|x,0\rangle$ and $|g,1\rangle$. Only in the regime  of {\em strong atom-cavity coupling}, with $g_\mathrm{0}\gg\{\kappa,\gamma\}$, the damping is weak enough so that vacuum-Rabi oscillations do occur. The other extreme is the {\em bad-cavity regime}, with $\kappa\gg g_\mathrm{0}^2/\kappa \gg \gamma$, which results in strong damping and quasi-stationary quantum states of the coupled system if it is continuously driven.

Two properties of the cavity can be used to distinguish between these regimes: First the strength of the atom-cavity coupling, $g_\mathrm{0}\propto 1/\sqrt{V}$ (dependant upon the mode volume of the cavity), and second the finesse $\mathcal{F}=\pi\sqrt{R}/(1-R)$ of the resonator, which depends on the mirror reflectivity $R$. The finesse gives the mean number of round trips in the cavity before a photon is lost by transmission through one of the cavity mirrors, and it is also identical to the ratio of free spectral range $\Delta\omega_\mathrm{FSR}$ to cavity linewidth $2 \kappa$. To reach strong coupling, a high value of $g_\mathrm{0}$ and therefore a short cavity of small mode volume are normally required. Keeping $\kappa$ small enough at the same time then calls for a high finesse and a mirror reflectivity $R \ge 99.999\%$.   

\subsection{Single-photon emission}
\label{sec:single-photon}
For the deterministic generation of single photons from coupled atom-cavity systems, all schemes implemented to date rely on the \keyword{Purcell effect} \cite{Purcell46}. The spatial mode density in the cavity and the coupling to the relevant modes is substantially different from free space \cite{Kleppner81}, such that the spontaneous photon emission into a resonant cavity gets either enhanced (${f > 1}$) or inhibited (${f < 1}$) by the Purcell factor 
\[f =   \frac{3Q\lambda^3}{4\pi^2V},\]
depending on the cavity's mode volume, $V$, and quality factor, $Q$. More importantly, the probability of spontaneous emission placing a photon into the cavity is given by  $\beta = f/(f+1)$.  If the mode volume of the cavity is sufficiently small, the emitter and cavity couple so strongly that $\beta\approx 1$, i.e. emissions into the cavity outweigh spontaneous emissions into free space. A deterministic photon emission into a single field mode is therefore possible with an efficiency close to unity. These effects have first been observed by Carmichael et al. \cite{Carmichael85} and  De Martini et al. \cite{Martini87}. Moreover, with the coherence properties uniquely determined by the parameters of the cavity and the driving process, one should be able to obtain indistinguishable photons from different cavities.  Furthermore the reversibility of the photon generation process, and quantum networking between different cavities has been predicted \cite{Cirac97,DiVincenzo00,Dilley12}, and demonstrated \cite{Boozer07,Mucke10,Specht11}.
We now introduce  different ways of producing single photons from such a system. These include \keyword{cavity-enhanced spontaneous emission} and Raman transitions stimulated by the vacuum field while driven by classical laser pulses. In particular, we discuss a scheme for \keyword{adiabatic coupling} between a single atom and an optical cavity, which is based on a unitary evolution of the coupled atom-cavity system \cite{Kuhn99,Vitanov01}, and is therefore intrinsically reversible. 

For a photon emission from the cavity to take place, it is evident that a finite value of $\kappa$ is mandatory, otherwise any light would remain trapped between the mirrors. Moreover, as $\kappa$ is the decay rate of the cavity field, the associated  duration of an emitted photon is typically $\kappa^{-1}$ or more. We also emphasise that $\gamma$ plays a crucial role in most experimental settings, since it accounts for the spontaneous emission into non-cavity modes, and therefore leads to a reduction of efficiency. The relation of the \keyword{atom-cavity coupling} constant and the Rabi frequency of the driving field to the two decay rates can be used for marking the difference between three basic classes of single-photon emission schemes from cavity-QED systems.\\
 
\begin{figure}
	\centering\includegraphics[width=4.4in]{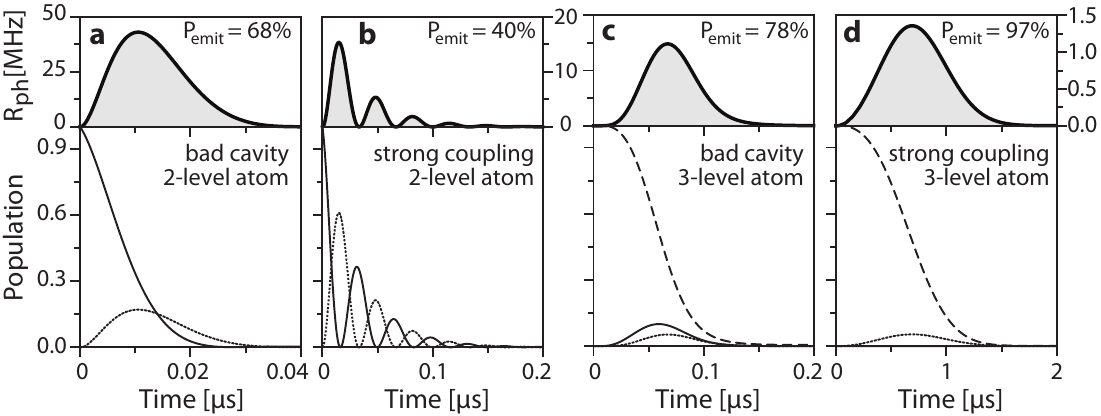}
	\caption{Evolution of the atomic states and photon emission rate $R_\mathrm{ph}=2\kappa\rho_\mathrm{gg}$ in different coupling cases (adapted from \cite{Solomon13}):  (a) and (b) are for an excited two-level atom  coupled to the cavity, showing populations $\rho_\mathrm{xx}$ (solid) and $\rho_\mathrm{gg}$ (dotted) of the product states $|x,0\rangle$ and $|g,1\rangle$. (c) and (d) are for a three-level atom-cavity system prepared in $|e,0\rangle$ and exposed to a pump pulse driving $|e\rangle - |x\rangle$ while the cavity couples $|x\rangle$ and $|g\rangle$. The initial-state population $\rho_\mathrm{ee}$  is dashed. (a) and (c) display the bad-cavity regime with $(g_\mathrm{0},\gamma,\kappa)=2\pi\times(15, 3, 20)\,$MHz, while (b) and (d) depict the strong-coupling case with $(g_\mathrm{0},\gamma,\kappa)=2\pi\times(15, 3, 2)\,$MHz. The pump pulses read $\Omega(t)=g_\mathrm{0} \sin(\pi   t/200\,ns)$ in (c), and $\Omega(t)=g_\mathrm{0} \times t/(1\,\mu$s) in (d). No transient population is found in $\rho_\mathrm{xx}$ in the latter case. The overall photon-emission probability reads always  $P_\mathrm{Emit}=\int R_\mathrm{ph}dt$.
	\label{fig:purcellregimes}}
\end{figure}

\noindent{\bf Cavity-enhanced spontaneous emission:}
We assume that a sudden excitation process (e.g. a short $\pi$ pulse with $\Omega\gg\{g_\mathrm{0}, \kappa, \gamma\}$, or some internal relaxation cascade from energetically higher states) drives the atom suddenly into its excited state $|x,0\rangle$. From there, a photon gets spontaneously emitted   either into the cavity or into free space. To analyse the process, we simply consider an excited two-level atom coupled to an empty cavity. This particular situation is the textbook example of cavity-QED that has been thoroughly analysed in the past. In fact, it has been proposed by Purcell \cite{Purcell46} and demonstrated by Heinzen et al. \cite{Heinzen87} and Morin et al. \cite{Morin94} that the spontaneous emission properties of an atom coupled to a cavity are significantly different from those in free space. For an analysis of the atom's behaviour, it suffices to look at the evolution of the $n=1$ Jaynes-Cummings doublet under the influence of the atomic polarisation decay rate $\gamma$ and the cavity-field decay rate $\kappa$. Non-cavity spontaneous decay of the atom and photon emission through one of the cavity mirrors both lead the system into state $|g,0\rangle$, which does not belong to the $n=1$ doublet. Therefore we can deal with these decay processes phenomenologically by introducing non-hermitian damping terms into the interaction Hamiltonian,
\begin{equation}
    H'_\mathrm{int}= - \hbar g_\mathrm{0}
    \left(|x\rangle\langle g| \hat{a} + \hat{a}^\dag |g\rangle\langle x|\right)
    -i \hbar \gamma |x\rangle\langle x| - i \hbar \kappa \hat{a}^\dag \hat{a}.
\end{equation}
Fig.\,\ref{fig:purcellregimes}(a) shows the time evolution of the atom-cavity system when $\kappa> g_\mathrm{0}$. The strong damping of the cavity  inhibits any vacuum-Rabi oscillation, since the photon is emitted from the cavity before it can be reabsorbed by the atom.  Therefore the transient population in state $|g,1\rangle$ is negligible and the adiabatic approximation
$\dot{c}_\mathrm{g}\approx 0$ can be applied, which gives
\begin{equation}
    \frac{d}{dt}c_\mathrm{x}=-\gamma c_\mathrm{x} -\frac{g_\mathrm{0}^2}{\kappa} c_\mathrm{x},
\end{equation}
with the solution
\begin{equation}
    c_\mathrm{x}(t)=\exp\left(-\left[\gamma+\frac{g_\mathrm{0}^2}{\kappa}\right]t\right).
\end{equation}
It is straightforward to see that the ratio of the emission rate into the cavity, $g_\mathrm{0}^2/\kappa$, to the spontaneous emission probability into free space becomes $g_\mathrm{0}^2/(\kappa\gamma)\equiv f$, i.e. the \keyword{Purcell factor}. It equals twice the one-atom cooperativity parameter, $C$, originally introduced in the context of optical bistability \cite{Lugiato84}. Hence the photon-emission probability from the cavity reads $P_\mathrm{\mbox{\tiny Emit}}=2C/(2C+1)$. Note that the atom radiates mainly into the cavity if $g^2_\mathrm{0}/\kappa\gg\gamma$. Together with $\kappa\gg g_\mathrm{0}$, this condition constitutes the \keyword{bad-cavity regime}.

The other extreme is  \keyword{strong coupling}, with $g_\mathrm{0}\gg(\kappa,\gamma)$. In this case vacuum Rabi oscillations between $|x,0\rangle$ and $|g,1\rangle$ occur, with both states decaying at the respective rates $\gamma$ and $\kappa$. Fig.\,\ref{fig:purcellregimes}(b) shows a situation where the atom-cavity coupling, $g_\mathrm{0}$, saturates the $|x,0\rangle \leftrightarrow |g,1\rangle$ transition. On average, the probabilities to find the system in either one of these two states are equal, and therefore the average ratio of the emission probability into the cavity to the spontaneous emission probability into free space is given by $\kappa/\gamma$. The \keyword{vacuum-Rabi oscillation} gives rise to an amplitude modulation of the photons emitted from the cavity here.\\

\noindent{\bf Bad-cavity regime:}
To take the effect of a slow excitation process into account, we  consider a $\Lambda$-type three-level atom coupled to a cavity. In the bad-cavity regime, $\kappa\gg g_\mathrm{0}^2/\kappa \gg \gamma$, the loss of excitation into unwanted modes of the radiation field is small and we may follow Law et al. \cite{Law96,Law97}.  We assume that the atom's $|e\rangle-|x\rangle$ transition is excited by a pump laser pulse while the atom emits a photon into the cavity by enhanced spontaneous emission.  The cavity-field decay rate $\kappa$ sets the fastest time scale, while the spontaneous emission rate into the cavity, $g_\mathrm{0}^2/\kappa$, dominates the incoherent decay of the polarisation from the excited atomic state. Provided any decay leads to a loss from the three-level system, the evolution of the wave vector is governed by the non-Hermitian Hamiltonian
\begin{equation}
    H'_\mathrm{int}=H_\mathrm{int}-i\hbar\kappa \hat{a}^\dag \hat{a}- i\hbar \gamma |x\rangle\langle x|,
\end{equation}
with $H_\mathrm{int}$ from Eq.\,(\ref{eq:Hatcav3}). To simplify the analysis, we take only the vacuum state, $|0\rangle$, and the one-photon state, $|1\rangle$, into account, thus that the state vector reads
\begin{equation}
    |\Psi(t)\rangle = c_\mathrm{e}(t)|e,0\rangle + c_\mathrm{x}(t)|x,0\rangle + c_\mathrm{g}(t)|g,1\rangle, 
\end{equation}
where $c_\mathrm{e}$, $c_\mathrm{x}$ and $c_\mathrm{g}$ are complex amplitudes. Their time evolution is given by the Schr\"{o}dinger equation, $i\hbar\frac{d}{dt}|\Psi\rangle = H'_\mathrm{int} |\Psi\rangle$, which yields
\begin{eqnarray}
    i\dot{c}_\mathrm{e} &=& \mbox{$\frac{1}{2}$} \Omega(t) c_\mathrm{x} \nonumber\\
    i\dot{c}_\mathrm{x} &=& \mbox{$\frac{1}{2}$} \Omega(t) c_\mathrm{e} + g_\mathrm{0} c_\mathrm{g} - i \gamma  c_\mathrm{x}\label{eq:cucecg}\\
    i\dot{c}_\mathrm{g} &=& g_\mathrm{0} c_\mathrm{x} - i \kappa c_\mathrm{g},\nonumber
\end{eqnarray}
with the initial condition $c_\mathrm{e}(0)=1$, $c_\mathrm{x}(0)=c_\mathrm{g}(0)=0$ and $\Omega(0)=0$. An adiabatic solution of (\ref{eq:cucecg}) is found if the decay is so fast that $c_\mathrm{x}$ and $c_\mathrm{g}$ are nearly time independent. This allows one to make the approximations $\dot{c}_\mathrm{x}=0$ and $\dot{c}_\mathrm{g}=0$, with the result
\begin{eqnarray}
    c_\mathrm{e}(t) &\approx& \exp\left(-\frac{\alpha}{4}\int_\mathrm{0}^t \Omega^2(t') dt' \right)\nonumber\\
    c_\mathrm{x}(t) &\approx& - \mbox{$\frac{i}{2}$} \alpha \Omega(t) 
    c_\mathrm{e}(t)    \label{eq:adsol}\\
    c_\mathrm{g}(t)&\approx& - \mbox{$\frac{i}{\kappa}$} g_\mathrm{0} c_\mathrm{x}(t),\nonumber
\end{eqnarray}
where $\alpha = 2 / (2\gamma +2g_\mathrm{0}^2/\kappa)$. Photon emissions from the cavity occur if the system is in $|g,1\rangle$, at the photon-emission rate $R_\mathrm{ph}(t) = 2\kappa |c_\mathrm{g}(t)|^2$. This yields a photon-emission probability of
\begin{eqnarray}
    P_\mathrm{Emit}&=&\int R_\mathrm{ph}(t) dt    \label{eq:pemission}\\
    &=&\frac{g_\mathrm{0}^2\alpha}{\kappa}\left[1-\exp\left(-\frac{\alpha}{2}\int \Omega^2(t) dt\right)\right]
    \longrightarrow \frac{g_\mathrm{0}^2\alpha}{\kappa}.\nonumber
\end{eqnarray}
Note that the exponential in Eq.\,(\ref{eq:pemission}) vanishes if the area $\int \Omega(t) dt$ of the exciting pump pulse is large enough. In this limit, the overall photon-emission probability does not depend on the shape and amplitude of the pump pulse. With a suitable choice of $g_\mathrm{0}$, $\alpha$, and $\kappa$, high photon-emission probabilities can be reached \cite{Law97}. Furthermore, as the stationary state of the coupled system depends on $\Omega(t)$, the  time envelope of the photon can be controlled to a large extent. \\

\noindent{\bf Strong-coupling regime:}
\label{sec:adipass}
To study the effect of the exciting laser pulse in the strong-coupling regime, we again consider a $\Lambda$-type three-level atom coupled to a cavity. We assume that the strong-coupling condition also applies to the Rabi frequency of the driving field, i.e. $\{g_\mathrm{0}, \Omega\}\gg\{\kappa, \gamma\}$. In this case, we can safely neglect the effect of the two damping rates on the time scale of the excitation. We then seek a method to effectively stimulate a Raman transition between the two ground states that also places a  photon into the cavity. For instance, the driving process can be   implemented in form of an adiabatic passage (STIRAP process \cite{Kuhn99,Vitanov01}) or a far-off resonant Raman process to   avoid  any transient population of the excited state, thus   reducing losses due to spontaneous emission into free space. An efficiency for photon generation close to unity can be   reached this way. Once a photon is placed into the cavity, it gets emitted   due to the finite cavity lifetime. 

The most promising approach is to implement an adiabatic passage in the optical domain between the two ground states \cite{Kuhn02:2,Kuhn03}. In fact, adiabatic passage methods have been used for coherent population transfer in atoms or molecules for many years. For instance, if a Raman transition is driven by two distinct pulses of variable amplitudes, effects like electromagnetically induced transparency (EIT) \cite{Harris93,Harris97}, slow light \cite{Hau99,Phillips01}, and stimulated Raman scattering by adiabatic passage (\keyword{STIRAP}) \cite{Vitanov01} are observed. These effects have been demonstrated with classical light fields and have the property in common that the system's state vector, $|\Psi\rangle$, follows a dark eigenstate, e.g. $|\phi^0_\mathrm{n}\rangle$, of the time-dependent interaction Hamiltonian. In principle, the time evolution of the system is completely controlled by the variation of this eigenstate. However, a more detailed analysis \cite{Kuhn02:2,Messiah59} reveals that the eigenstates must change slowly enough to allow adiabatic following. Only if this condition is met, a three-level atom-cavity system, once prepared in $|\phi^0_\mathrm{n}\rangle$,  stays there forever, thus allowing one to control the relative population of  $|e,n-1\rangle$ and $|g,n\rangle$ by  adjusting the pump Rabi frequency $\Omega$. This is obvious for a system initially prepared in  $|e,n-1\rangle$. As can be seen from Eq.\,(\ref{anull1}), that state coincides with $|\phi^0_\mathrm{n}\rangle$ if the condition $2 g_\mathrm{0}\sqrt{n}\gg\Omega$ is initially met. Once the system has been prepared in the dark state, the ratio between the populations of the contributing states reads
\begin{equation}
    \frac{|\langle e,n-1|\Psi\rangle|^2}{|\langle g,n|\Psi\rangle|^2} =
    \frac{4 n g_\mathrm{0}^2}{\Omega^2}.
\end{equation}
As proposed in \cite{Kuhn99}, we  assume that an atom in state $|e\rangle$ is placed into an empty cavity, which nonetheless drives the $|g,1\rangle\leftrightarrow|x,0\rangle$ transition with the Vacuum Rabi frequency $2g_\mathrm{0}$. The initial state $|e,0\rangle$ therefore coincides with $|\phi^0_\mathrm{1}\rangle$ as long as no pump laser is applied. The atom then gets exposed to a laser pulse coupling the $|e\rangle\leftrightarrow|x\rangle$ transition with a slowly rising amplitude that  leads to $\Omega\gg 2 g_\mathrm{0}$. In turn, the  atom-cavity system evolves from $|e,0\rangle$ to $|g,1\rangle$, thus increasing the photon number by one. This  scheme can be seen as vacuum-stimulated Raman scattering by adiabatic passage, also known as \keyword{V-STIRAP}.  If we assume a  cavity decay time, $\kappa^{-1}$ much longer than the interaction time, a photon is emitted from the cavity with a probability close to unity and with properties uniquely defined by $\kappa$, after the system has been excited to $|g,1\rangle$. 

In contrast to such an idealised scenario, Fig.\,\ref{fig:purcellregimes}(d) shows a more realistic situation where a photon is generated and already emitted from the cavity during the excitation process.  This is due to the cavity decay time being comparable or shorter than the duration of the exciting laser pulse. Even in this case, no secondary excitations or photon emissions can take place. The system eventually reaches the decoupled state $|g,0\rangle$ when the photon escapes. However, the photon-emission probability is slightly reduced as the non-Hermitian contribution of $\kappa$ to the interaction Hamiltonian is affecting the dark eigenstate $|\phi^0_\mathrm{1}\rangle$ of the Jaynes-Cummings triplet (\ref{anull1}). It now has a small admixture of $|x,0\rangle$ and hence is weakly affected by spontaneous emission losses \cite{Kuhn02:2}. 

\subsubsection{Single-photon emission from atoms or ions in cavities}
Many revolutionary photon generation schemes have recently been demonstrated, such as a single-photon turnstile device based on the Coulomb blockade mechanism in a quantum dot \cite{Kim99}, the fluorescence of a single molecule \cite{Brunel99,Lounis00}, or a single colour centre (Nitrogen vacancy) in diamond \cite{Kurtsiefer00,Brouri00}, or the photon emission of a single quantum dot into free space \cite{Michler00,Santori01,Yuan02}.  All these  schemes emit photons upon an external trigger event. The photons are spontaneously emitted into various modes of the radiation field, e.g. into all directions,  and often show a  broad energy distribution.  
Nonetheless these photons are excellent for quantum cryptography and communication, and also a reversal of the free-space spontaneous emission has been demonstrated, see chapters by Leuchs \& Sondermann, Chuu \& Du, and Piro \& Eschner. Cavity QED has the potential of performing this bi-directional state mapping between atoms and photons very effectively,  and thus is expected to levy many fundamental limitations to scalability in quantum computing or quantum networking. We therefore  focus here on cavity-enhanced emission techniques into well-defined modes of the radiation field.\\
   
\begin{figure}[h]
 \centering\includegraphics[width=4.4in]{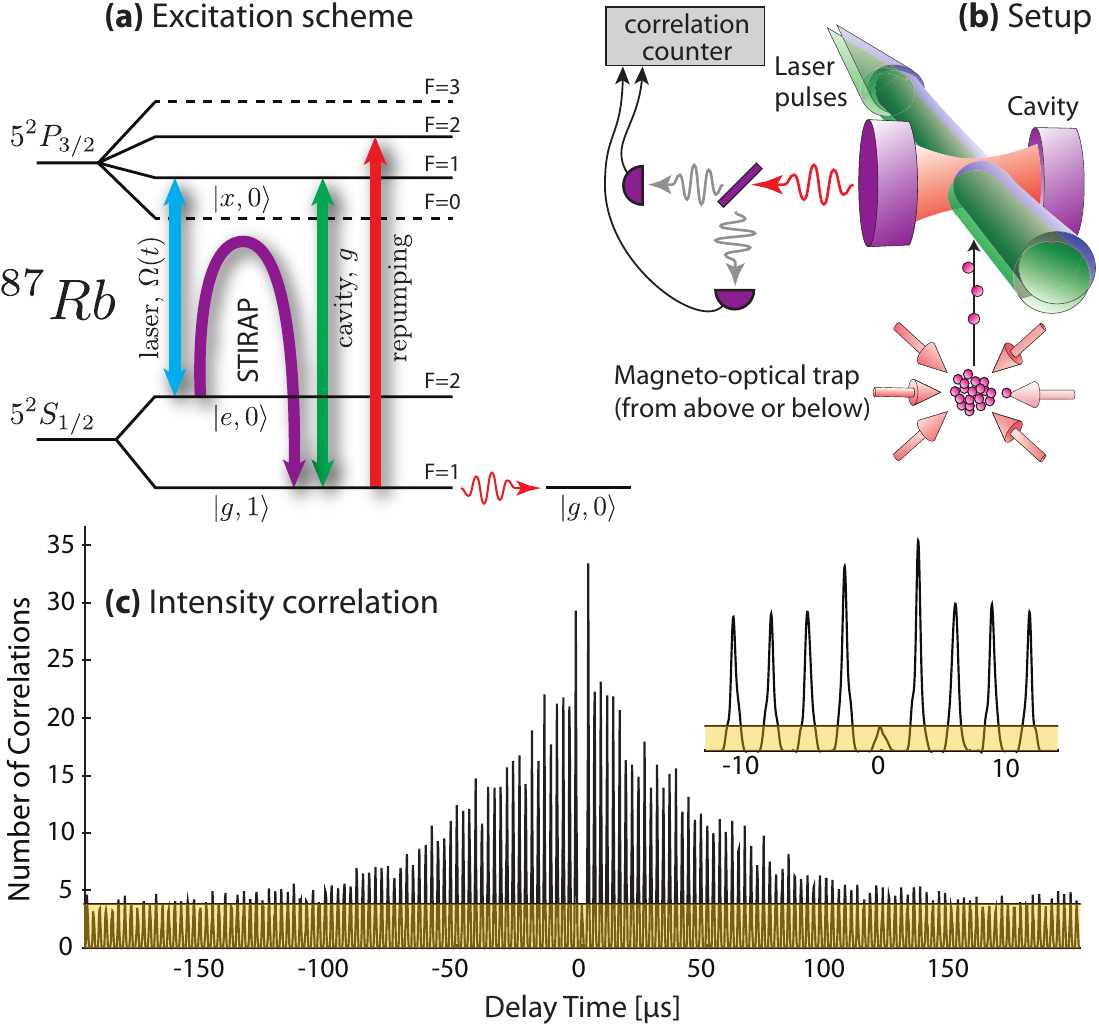}
 \caption{Single-photon source based on atoms travelling through an optical cavity. \textbf{(a)} Excitation scheme realised in $^{87}$Rb for the pulsed single-photon generation. The atomic states labeled $\left| e\right\rangle$, $\left| x\right\rangle $ and $\left|g\right\rangle $ are involved in the Raman process, and the states $\left| 0\right\rangle $ and $\left| 1\right\rangle $ denote the photon number in the cavity. \textbf{(b)} A cloud of laser-cooled atoms  moves through an optical cavity either from above \cite{Kuhn02}, or from below using an atomic fountain \cite{Nisbet11}. Laser pulses travel perpendicular to the cavity axis to control the emission process. The light is analysed using a \keyword{Hanbury-Brown \& Twiss} (\keyword{HBT}) setup with a pair of avalanche photodiodes. \textbf{(c)} Intensity correlation of the emitted light measured with the HBT setup, with atoms injected using an atomic fountain \cite{Nisbet11}. The  contribution of correlations between real photons and detector dark counts is shown in yellow. \label{fig:falling_atoms}} 
\end{figure}

\noindent{\bf Neutral  atoms:}
A straightforward implementation of a cavity-based single-photon source consists of a single atom placed between two  cavity mirrors, with a stream of laser pulses travelling perpendicular to the cavity axis to trigger photon emissions. The most simplistic approach to achieve this is by sending a dilute atomic beam through the cavity, with an average number of atoms in the mode far below one. However, for a thermal beam, the obvious drawback would be an interaction time between atom and cavity far too short to achieve any control of the exact photon emission time. Hence cold (and therefore slow) atoms are required to overcome this limitation. The author followed this route \cite{Kuhn02,Nisbet11}, using a magneto-optical trap to cool a cloud of rubidium atoms below $100\,\mu$K at a distance close to the cavity. Atoms released from the trap eventually travel through the cavity, either falling from above or being injected from below in an atomic fountain. Atoms enter the cavity randomly, but interact with its mode for $20-200\,\mu$s. Within this limited interaction time, between 20 and 200 single-photon emissions can be triggered. Fig.\,\ref{fig:falling_atoms} illustrates this setup, together with the excitation scheme between hyperfine states in $^{87}$Rb used to generate single photons by the adiabatic passage technique discussed on page \pageref{sec:adipass}. 

Bursts of single photons are emitted from the cavity whenever a single atom passes its mode, and strong antibunching is found in the photon statistics, as shown in Fig.\,\ref{fig:falling_atoms}(a). A \keyword{sub-Poissonian} \keyword{photon statistics} is found when conditioning the experiment on the actual presence of an atom in the cavity \cite{Hennrich04}.  In many cases, this is automatically granted -- a good example is the characterisation  of the photons by \keyword{two-photon interference}. For such experiments, pairs of photons are needed that meet simultaneously at a beamsplitter. With just one source under investigation, this is achieved with a long optical fibre delaying the first photon of a pair of successively emitted ones. With the occurrence of such photon pairs being the precondition to observe any correlation and the probability for successive photon emissions being vanishingly small without atoms, the presence of an atom is actually assured whenever data is recorded. 

\begin{figure}[t]
\centering\includegraphics[width=4.4in]{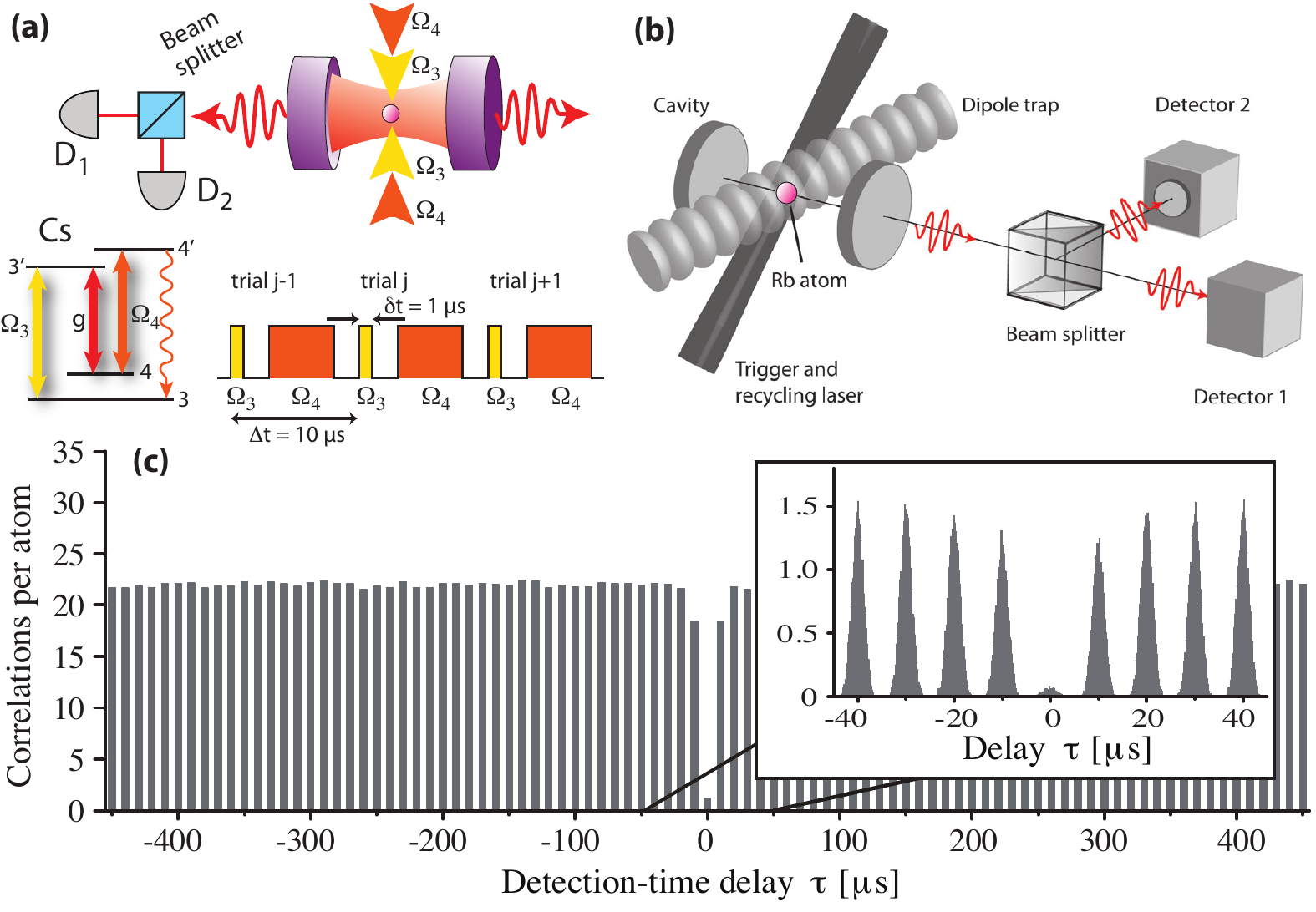}
\caption{Atom-cavity systems with a single atom at rest in the cavity mode. 
{\bf (a):} The setup by J. McKeever et al. \cite{McKeever04} is using a dipole trap running along the cavity axis to hold a single caesium atom in the cavity. The cavity is symmetric, so that half the photons are directed towards a pair of detectors for analysing the photon statistics.  
{\bf (b):} The author has been using a dipole trap running perpendicular to the cavity axis, The trap is holding a single rubidium atom in the cavity \cite{Hijlkema07}. The cavity is asymmetric, and  photons emitted through its output coupler are directed to a pair of photon counters to record the second-order correlation function of the photon stream. 
In both cases, the trapped atom is exposed to a sequence of laser pulses that trigger the photon emission, cool the atom, and re-establish the initial condition by optical pumping.
{\bf (c):} Intensity correlation function of the light emitted form a trapped-atom-cavity system, as found by the author \cite{Hijlkema07}. 
\label{fig:atomtrap}}
\end{figure}

Only lately, refined versions of this type of photon emitter have been realised, with a single atom held in the cavity using a dipole-force trap. McKeever et al. \cite{McKeever04} managed to hold a single Cs atom in the cavity with a dipole-trapping beam running along the cavity axis, while Hijlkema et al.  \cite{Hijlkema07} are using a combination of dipole trapping beams running perpendicular and along the cavity to catch and hold a single Rb atom in the cavity mode. As illustrated in Fig.\,\ref{fig:atomtrap}, the trapped atom is in both cases exposed to a sequence of laser pulses alternating between triggering the photon emission, cooling and repumping the atom to its initial state to repeat the sequence. The atom is trapped, so that the photon statistics is not affected by fluctuations in the atom number and therefore is sub-Poissonian, see Fig.\,\ref{fig:atomtrap}(c). Moreover, with trapping times for single atoms up to a minute, a quasi-continuous bit-stream of photons is obtained.  

The major advantage of using neutral atoms as photon emitters in Fabry-Perot type cavities is that a relatively short cavity (some 100\,$\mu$m) of high finesse (between $10^5$ and $10^6$) can be used. One thus obtains strong atom-cavity coupling, and the photon generation can be driven either in the steady-state regime or dynamically by V-STIRAP. This allows one to control the coherence properties and the shape of the photons to a large extent, as discussed in section \ref{sec:shape}. Photon generation efficiencies as high as 65\% have been reported with these systems. Furthermore, based on the excellent coherence properties, first applications such as atom-photon entanglement and atom-photon state mapping \cite{Wilk07-Science, Weber09,Ritter12,Nolleke13} have recently been demonstrated.  

Apart from the above Fabry-Perot type cavities, many other micro-structured cavities have been explored during the last years. These often provide a much smaller mode volume and hence boost the atom-cavity coupling strength by about an order of magnitude. However, this goes hand-in-hand with increased cavity losses and thus a much larger cavity linewidth, which might be in conflict with the desired addressing of individual atomic transitions. Among the most relevant new developments are fibre-tip cavities, which use dielectric Bragg stacks at the tip of an optical fibre as cavity mirrors \cite{Colombe07, Trupke07}.  Due to the small diameter of the fibre, either two fibre tips can be brought very close together, or a single fibre tip can be complemented  by a micro-structured mirror on a chip to form a high-finesse optical cavity. A slightly different approach is the use of ring-cavities realised in solid state, guiding the light in a whispering gallery mode. An atom can be easily coupled to the evanescent field of the cavity mode, provided it can be brought close to the surface of the substrate. Nice examples are microtoroidal cavities realised at the California Institute of Technology \cite{Dayan08,Aoki09}, and bottle-neck cavities in optical fibres \cite{Pollinger09}. These cavities have no well-defined mirrors and therefore no output coupler, so one usually arranges for emission into well-defined spatio-temporal modes via evanescent-field coupling to the core of an optical fibre.

We would like to remind the reader at this point that a large variety of other exciting cavity-QED experiments have been performed that were not aiming at a single-photon emission in the optical regime. Most important amongst those are the coupling of Rydberg atoms \cite{Rauschenbeutel00} or superconductive SQUIDs \cite{Wallraff04} to microwave cavities, which is also a well-established way of placing single photons into a cavity using either $\pi$ pulses \cite{Maitre97} or dark resonances \cite{Brattke01}. Also the coupling of ultracold quantum gases to optical cavities has been studied extensively \cite{Brennecke07,Colombe07}, and has proven to be a useful method to acquire information on the atom statistics. Last but not least, large efforts have been made to study cavity-mediated forces on either single atoms or atomic ensembles \cite{Vuletic00,Vuletic01,McKeever03,Maunz04,Thompson06, Fortier07}, which  lead to the development of cavity-mediated cooling techniques.\\

\begin{figure}[t]
\centering\includegraphics[width=4.4in]{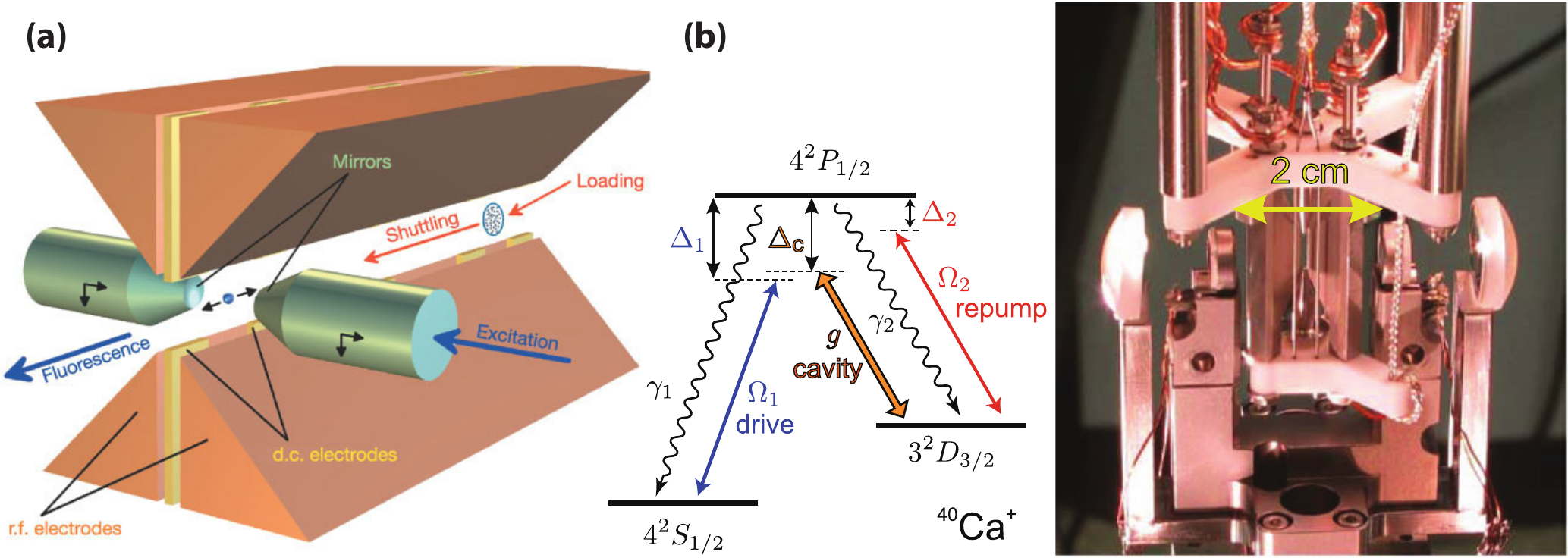}
\caption{Arrangement of ion-trap electrodes and cavity in \textbf{(a)} the experiment by M.\,Keller et al. \cite{Keller04}. The ion is shuttled to the cavity region after loading. Upon excitation of the ion from the side of the cavity, a single photon gets emitted into the cavity mode
(Reprinted by permission from Nature Publishing Group: Nature, G.\,Guth\"ohrlein et al. \cite{Guthoerlein01}, Copyright 2001).
The ion-cavity arrangement and excitation scheme in $^{40}$Ca$^+$ studied by C.\,Russo and H.\,G.\,Barros et al. \cite{Russo09,Barros09} in Innsbruck \textbf{(b)} is using a near-concentric cavity which leads to an increased density of otherwise non-degenerate transverse modes
(Panel b adapted with permission from Springer: Applied Physics B, Russo et al. \cite{Russo09}, Copyright 2009).
\label{fig:Lange}}
\end{figure}

\noindent{\bf Trapped ions:}
Although neutral-atom systems have their advantages for the generation of single photons, such experiments are sometimes subject to undesired variations in the atom-cavity coupling strength and multi-atom effects. Also trapping times are still limited in the intra-cavity dipole-trapping of single atoms. A possible solution is to use a strongly localised single ion in an optical cavity, as has first been demonstrated by M. Keller et al. \cite{Keller04}. In their experiment, an ion is optimally coupled to a well-defined field mode, resulting in the reproducible generation of single-photon pulses with precisely defined timing.  The stream of emitted photons is uninterrupted over the storage time of the ion, which, in principle, could last for several days.

The major difficulty in combining an ion trap with a high-finesse optical cavity comes from the dielectric cavity mirrors, which influence the trapping potential if they get too close to the ion. This effect might be detrimental in case the mirrors get electrically charged during loading of  the ion trap, e.g. by the electron beam used to ionise the atoms. Fig.\,\ref{fig:Lange}(a) shows how this problem has been solved in \cite{Keller04} by shuttling the trapped ion from a spatially separate loading region into the cavity. Nonetheless, the cavity in these experiments is typically more than 10--20\,mm long to avoid distortion of the trap. Thus the coupling to the cavity is weak, and although optimised pump pulses were used, the single-photon efficiency in \cite{Keller04} did not exceed  $(8.0 \pm 1.3)\%$. This is in good accordance with theoretical calculations, which also show that the efficiency can be substantially increased in future experiments by reducing the cavity length. It is important to point out that the low efficiency does not interfere with the singleness of the photons. Hence the $g^{(2)}$ correlation function of the emitted photon stream corresponds to the one depicted in Fig.\,\ref{fig:atomtrap}c, with $g^{(2)}(0)\rightarrow 0$. With an improved ion-cavity setup, H.\,G. Barros et al. \cite{Barros09} were able to reach a single-photon emission efficiency of $(88 \pm 17)\%$ in a cavity of comparable length, using a more favourable mode structure in the near-confocal cavity depicted in Fig.\,\ref{fig:Lange}(b) and far-off resonant Raman transitions between magnetic sublevels of the  ion.

\section{Cavity-enhanced atom-photon entanglement}
In their groundbreaking experiments, Chris Monroe \cite{Blinov04} and Harald Weinfurter \cite{Volz06}  successfully demonstrated the \keyword{entanglement} of the polarisation of a single photon with the spin of a single ion or atom, respectively. To do so, they drove a free-space excitation and emission scheme in a single trapped ion or atom that lead to two possible final spin states of the atom, $|\downarrow\rangle$ and $|\uparrow\rangle$ upon emission of either a $\sigma^+$ or $\sigma^-$ polarised photon, thus projecting the whole system into the entangled atom-photon state
\begin{equation}
(|\sigma^+,\downarrow\rangle-|\sigma^-,\uparrow\rangle)/\sqrt{2}.
\label{APentangle}
\end{equation}
Projective measurements on pairs of photons emitted from two distant atoms or ions were then used for entanglement swapping, thus resulting in the entanglement and teleportation of quantum states \cite{Olmschenk09}. Such photon-matter entanglement has a potential advantage of adding memory capabilities to quantum information protocols. In addition, this provides a quantum matter-light interface, thereby using different physical media for different purposes in a quantum information application. However, the spontaneous emission of photons into all directions is an inherent limitation of this approach. Even the best collection optics captures at most 25\% of the photons \cite{Bondo06}, with actual experiments reaching overall photon-detection efficiencies of about $5\times 10^{-4}$ \cite{Volz06}. Combined with the spontaneous character of the emission, efficiencies are very low and scaling is a serious issue. 

\begin{figure}[t]
  \centering\includegraphics[width=4.4in]{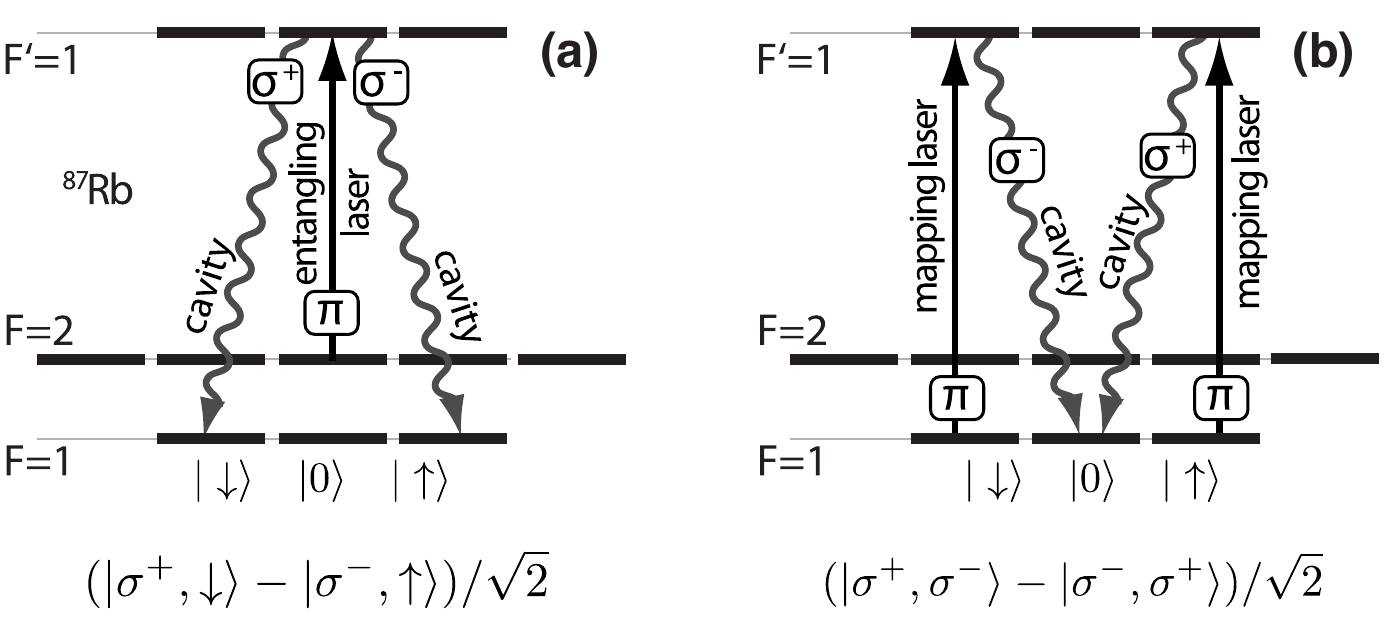}
    \caption{Entanglement and state mapping (from \cite{Wilk07-Science}): Laser pulses drive vacuum-stimulated Raman transitions, first (A) creating an entanglement between the atom and the emitted photon, and then (B) mapping the atomic state onto the polarization of a second photon. Entanglement is then shared between two flying photons, with the atom disentangled.\label{fig:entscheme}}
\end{figure}

We shall see in the following that the coupling of atoms or ions to optical cavities is one effective solution to this problem, with quantum state mapping and entanglement between atomic spin and photon polarisation recently achieved in cavity-based single-photon emitters \cite{Wilk07-Science,Weber09,Ritter12,Nolleke13}. Very much like what was discussed in the preceding sections, this is achieved using an intrinsically deterministic single-photon emission from a single $^{87}$Rb atom in strong coupling to an optical cavity \cite{Wilk07-Science}. The triggered emission of a first photon entangles the internal state of the atom and the polarization state of the photon. To probe the degree of entanglement,  the atomic state is then mapped onto the state of a second single photon.  As a result of the state mapping a pair of entangled photons is produced, one emitted after the other into the same spatial mode. The polarization state of the two photons is analyzed by tomography, which also probes the prior entanglement between the atom and the first photon. 

\begin{figure}[t]
\sidecaption
 \includegraphics[width=2.5in]{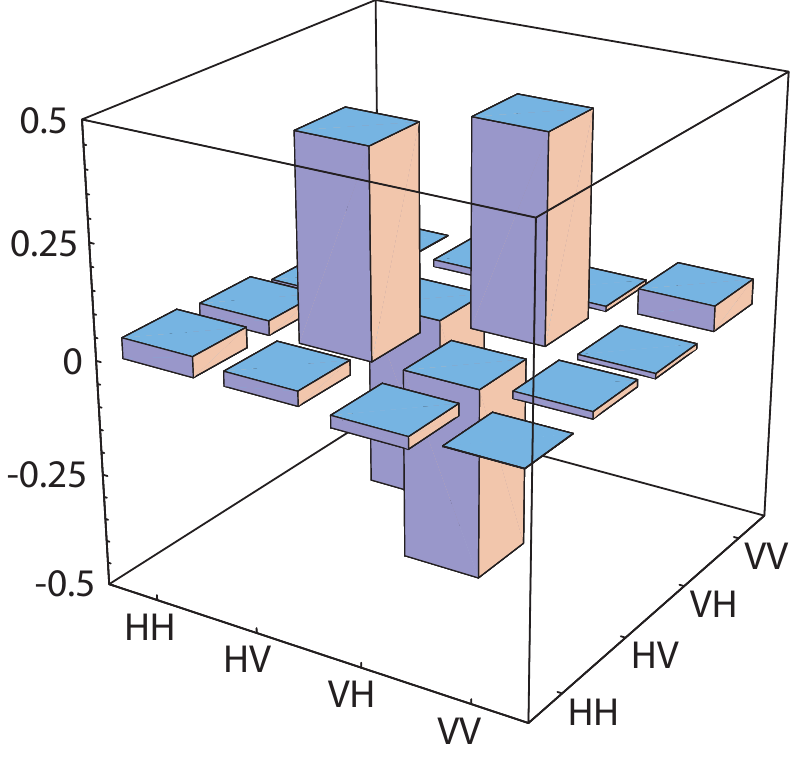}
    \caption{Real part of the density matrix reconstructed from quantum state tomography of successively emitted photon pairs (from \cite{Wilk07-Science}): All imaginary parts (not shown) have a magnitude smaller than 0.03. Fidelity with the expected Bell state is $86.0(4)\%$. About 2200 entanglement events were collected for each of nine measurement settings.\label{fig:dmatrix}}
\end{figure}

All the relevant steps of entanglement preparation and detection are shown schematically in Fig.\,\ref{fig:entscheme}. The rubidium atom is  prepared in the $|F=2, m_F=0\rangle$ state of the $5S_{1/2}$ ground level. Then a $\pi$-polarized laser pulse (polarised linearly along the cavity axis and resonant with the transition from $F=2$ to $F'=1$ in the excited $5P_{3/2}$ level) together with the cavity (coupling levels $F = 1$ and $F' = 1$) drives a Raman transition to the $|F = 1, m_F=\pm 1\rangle$ magnetic substates of the ground level. No other cavity-enhanced transitions are possible because the cavity only supports left- and right- handed circularly polarized $\sigma^+$ and $\sigma^-$ polarisation along its axis. Therefore the two different paths to $|\downarrow\rangle\equiv |F = 1, m_F = -1\rangle$ and $|\uparrow\rangle\equiv |F = 1, m_F = +1\rangle$ result in the generation of a $\sigma^+$ and $\sigma^-$  photon, respectively. Thus the atom becomes entangled with the photon and the resulting overall quantum state is identical to the one obtained in the free-space  experiments, outlined above in Eq.\,\ref{APentangle}. However, the substantial difference is that the photon gets deterministically emitted into a single spatial mode, well defined by the geometry of the surrounding cavity. Hence the success probability is close to unity, and a scalable arrangement of atom-cavity arrays is therefore in reach.

To probe the atom-photon entanglement established this way, the atomic state is mapped onto another photon in a second step. This photon can be easily analysed outside the cavity. To do so, a $\pi$-polarized laser pulse resonant with the transition from $F = 1$ to $F' = 1$ drives a second Raman transition in the cavity. This is transferring any atom from state $|\downarrow\rangle$ to $|0\rangle\equiv |F=1,m_F =0\rangle$ upon emission of a $\sigma^+$ photon, whereas $|\uparrow\rangle$ atoms are equally transferred to $|0\rangle$, but upon a  $\sigma^-$ emission. Hence the atom eventually gets disentangled, and a \keyword{polarization entanglement} 
\begin{equation}
(|\sigma^+,\sigma^-\rangle-|\sigma^-,\sigma^+\rangle)/\sqrt{2}
\label{PPentangle}
\end{equation}
is established between the two successively emitted photons. Because the photons are created in the same spatial mode, a non-polarizing beam splitter (NPBS) is used to direct the photons randomly to one of two measurement setups. This allows each photon to be detected in either the H/V, circular right/left (R/L), or linear diagonal/ antidiagonal (D/A) basis. Thus a full quantum state tomography is performed by measuring correlations between the photons in several different bases, selected by using different settings of half- and quarter-wave plates \cite{James01,Altepeter05}. These measurements lead to the reconstructed \keyword{density matrix} shown in Fig.\,\ref{fig:dmatrix}. It has only positive eigenvalues, and its fidelity with respect to the expected Bell state is $F = 86.0(4)\%$, with $0.5 < F \le 1$ proving entanglement.

Based on this successful atom-photon entanglement scheme, elementary \keyword{quantum-network} links implementing \keyword{teleportation} protocols between remote trapped atoms and atom-photon \keyword{quantum gate} operations have recently been demonstrated \cite{Weber09,Ritter12,Nolleke13,Reiserer14}. These achievements represent a big step towards a feasible quantum-computing system because they show how to overcome most scalability issues to quantum networking in a large distributed light-matter based approach.

\section{Photon coherence, amplitude and phase control}
The vast majority of single-photon applications not only rely on the deterministic emission of single photons, but also require them to be indistinguishable from one another. In other words, their mutual \keyword{coherence} is a key element whenever two or more photons are required simultaneously. The most prominent example to that respect is linear optics quantum computing (\keyword{LOQC}) as initially proposed by Knill, Laflamme and Milburn \cite{Knill01}, with its feasibility demonstrated using spontaneously emitted photon pairs from parametric down conversion \cite{OBrien03}. Scaling LOQC to useful dimensions calls for deterministically emitted indistinguishable photons. Furthermore, with photons used as information carriers, it is common practice to use their polarization, spatio-temporal mode structure or frequency for encoding classical or quantum state superpositions. To do so, the capability of shaping  photonic modes is essential. Several of these aspects are going to be discussed here. 

\begin{figure}[h]
 \centering\includegraphics[width=4.4in]{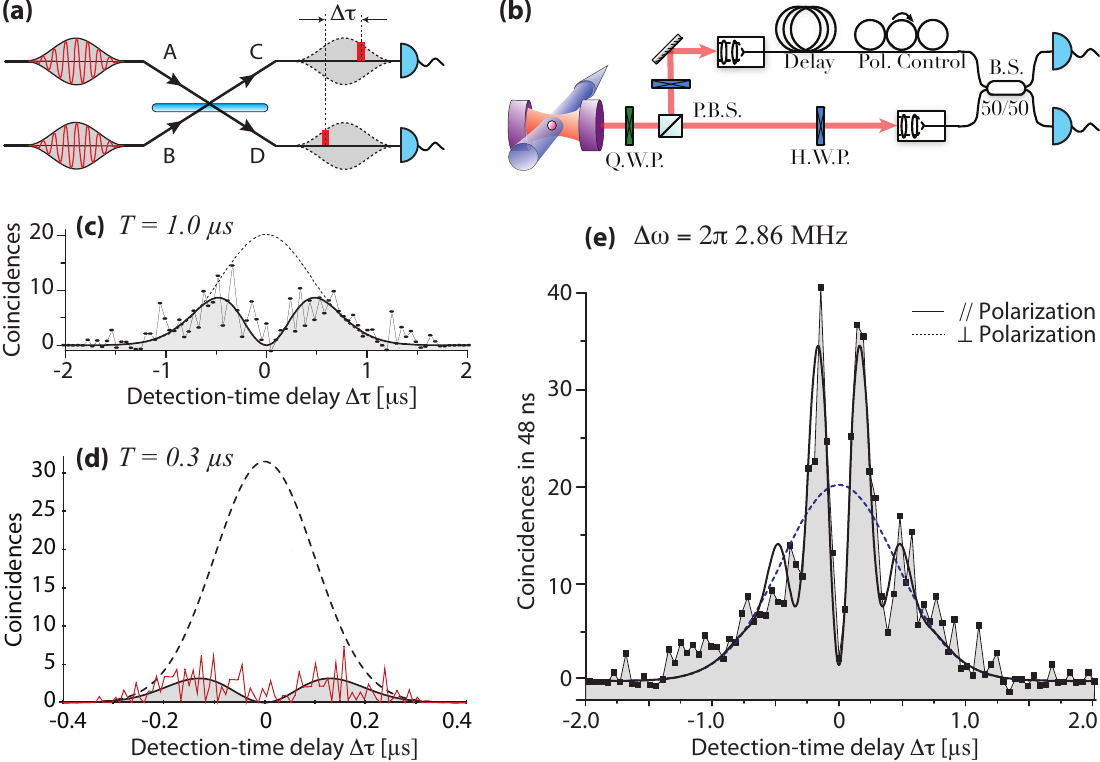}
 \caption{Time-resolved two-photon interference of photons arriving simultaneously at a beam splitter \textbf{(a)}. With photons emitted successively from one atom-cavity system, this has been achieved using an optical delay line \textbf{(b)}. Panel \textbf{(c)} shows the correlation function from \cite{Legero04} for photons of $1.0\,\mu$s duration as a function of detection-time delay $\Delta\tau$. A pronounced dip at the origin is found, with the dip-width indicating the photon coherence time. The dotted line shows correlations found if distinguishable photons of perpendicular polarisation are used, while the solid line depicts the correlations found if the photon polarisation is parallel. Panel \textbf{(d)} shows data from a more recent experiment \cite{Nisbet11} with photons of $0.3\,\mu$s duration. The photons are nearly indistinguishable and the integral two-photon coincidence probability drops to 20\% of the reference value found with non-interfering photons. Panel \textbf{(e)} shows data from \cite{Legero04,Legero06} with interfering photons of different frequency. This gives rise to a pronounced oscillation of the coincidence signal as a function of $\Delta\tau$ with the difference frequency $\Delta\omega$.
 \label{fig:HOM_resolved}}
\end{figure}

\subsection{Indistinguishability of photons}
At first glance, one would expect any single-photon emitter that is based on a single quantum system of well-defined level structure to deliver indistinguishable photons of well-defined energy. However, this is often not the case for a large number of reasons. For instance, multiple pathways leading to the desired single-photon emission or the degeneracy of spin states might lead to broadening of the spectral mode or to photons in different polarisation states, entangled with the atomic spin \cite{Wilk07-Science}. Also spontaneous relaxation cascades within the emitter result in a timing jitter of the last step of the cascade, which is linked to the desired photon emission. Nevertheless, atoms coupled to cavities have been shown to emit nearly indistinguishable photons with well defined timing. Their coherence properties are normally governed by the dynamics of the Raman process controlling the  generation of photons, and--surprisingly--not substantially limited by the properties and lifetime of the cavity mode or the atoms \cite{Legero04}.   

Probing photons for indistinguishability is often accomplished with a two-photon interference experiment of the \keyword{Hong-Ou-Mandel} (\keyword{HOM}) type.
For two identical photons that arrive simultaneously at different inputs of a 50:50 beam splitter, they bunch and then leave as a photon pair into either one or the other output port. Hence no correlations are found between two detectors that monitor the two outputs. This technique has been well established in connection with photons emitted from spontaneous parametric-down conversion (SPDC) sources, with the correlations between the outputs measured as a function of the {\em arrival-}time delay between photons. 

For the cavity-based emitters discussed here, the situation is substantially different. The bandwidth of these photons is very narrow, and therefore their coherence time (or length) might be extremely long, i.e. several $\mu$s (some 100$\,$m). The time resolution of the detectors is normally 3-4 orders of magnitude faster than this photon length, so that the two-photon correlation signal is now determined as a function of the detection-time delay, with the arrival-time delay of the long photons deliberately set to zero \cite{Legero03,Legero06}. This can be seen as a \keyword{quantum-homodyne} measurement at the single-photon level, with a single local-oscillator photon arriving at one port of a beam splitter, and a single signal photon arriving at the other port. To develop an understanding how the probability for photon coincidences between the two output ports of the beam splitter depends on the detection-time difference and the mutual phase coherence of the two photons, a step-by-step analysis of the associated quantum jumps and quantum state evolution is most instructive:

Prior to the first photodetection,  two photons arrive simultaneously in the input modes $A$ and $B$ at the beam splitter and the overall state of the system reads $|1_A1_B\rangle$. The first photo detection at time $t_\mathrm{1}$ in either output  $C$ or $D$ could have been of either photon, thus  the remaining quantum state  reduces to
\begin{equation}
|\psi(t_\mathrm{1})\rangle = (|1_A0_B\rangle\pm|0_A1_B\rangle)/\sqrt{2},
\end{equation}
where ``$+$" and ``$-$" correspond to the photodetection in port $C$ and $D$, respectively. We now assume that the second photodetection takes place $\Delta\tau$ later, at time $t_\mathrm{2}=t_\mathrm{1}+\Delta\tau$, with the input modes $A$ and $B$ having acquired a phase difference $\Delta\phi$ (for whatever reason) during that time interval. Hence prior to the second detection, the reduced quantum state has evolved to
\begin{equation}
|\psi(t_\mathrm{2})\rangle = (|1_A0_B\rangle\pm e^{i\Delta\phi}|0_A1_B\rangle)/\sqrt{2}.
\end{equation}
By consequence, the probability for the second photon being detected in the same port as the first photon is $P_\mathrm{same}=\cos^2(\Delta\phi/2)$, while the probability for the second photon being detected in the respective other beam-splitter port reads
\begin{equation}
P_\mathrm{other}=\sin^2(\Delta\phi/2).
\end{equation}
The probability $P_\mathrm{CD}$ for coincidence counts between the two detectors in the beam splitter's output ports $C$ and $D$  is therefore proportional to $\sin^2(\Delta\phi/2)$. This implies that any systematic variation of the phase difference $\Delta\phi$ between the two input modes $A$ and $B$ with time $\Delta\tau$ leads to a characteristic modulation of the coincidence function
\begin{equation}
g^{(2)}_\mathrm{CD}(\Delta\tau) = \frac{\langle P_C(t)P_D(t+\Delta\tau) \rangle_t}{\langle P_C\rangle\langle P_D\rangle} \propto \sin^2(\Delta\phi(\Delta\tau)/2).
\end{equation}
A good example is the analysis of two photons of different frequency. We consider one photon of well-defined frequency $\omega_\mathrm{0}$ acting as {\em local oscillator} arriving at port $A$ at the beam splitter, and another one of frequency $\omega_\mathrm{0}+\Delta\omega$ which we regard as {\em signal photon} arriving simultaneously at port $B$. Their mutual phase is undefined until the first photodetection at $t_\mathrm{1}$, and thereafter evolves according to $\Delta\phi(\Delta\tau) = \Delta\omega\times\Delta\tau$. In this case, the probability for coincidence counts between the beam splitter outputs,
\begin{equation}
P_\mathrm{CD}(\Delta\tau) \propto \sin^2(\Delta\omega\times\Delta\tau/2),
\end{equation}  
oscillates at the difference frequency between local oscillator and signal photon. This phenomenon has been extensively discussed in \cite{Legero04,Legero06} and is also illustrated in Fig.\,\ref{fig:HOM_resolved}. Furthermore, the figure  shows the effect of random \keyword{dephasing} on the time-resolved correlation function. For photons of 1\,$\mu$s duration, a 470\,ns wide dip has been found around $\Delta\tau=0$. Thereafter, the coincidence probability reached the same mean value that is found with non-interfering photons of, e.g., different polarisation. In this case, we  conclude that the dip-width in the coincidence function is identical to the mutual coherence time between the two photons. It is remarkable that it exceeds the decay time of both cavity and atom by one order of magnitude in that particular experiment. This proves that the photon's coherence is to a large extent controlled by the Raman process driving the photon generation, without being limited by the decay channels within the system. 

\subsection{Arbitrary shaping of amplitude and phase}
\label{sec:shape}
From the discussion in section \ref{sec:single-photon} we have seen that the dynamic evolution of the atomic quantum states determines the photon emission probability, and thereby also the photon's waveform. This raises the question as to what extent one can arbitrarily shape the photons in time by controlling the envelope of the driving field. This is  important  for applications such as quantum state mapping, where photon wave packets symmetric in space and time should allow for a time-reversal of the emission process \cite{Cirac97}. Employing photons of soliton-shape for dispersion-free propagation could also help boost quantum communication protocols.

\begin{figure}[t]
  \centering\includegraphics[width=4.4in]{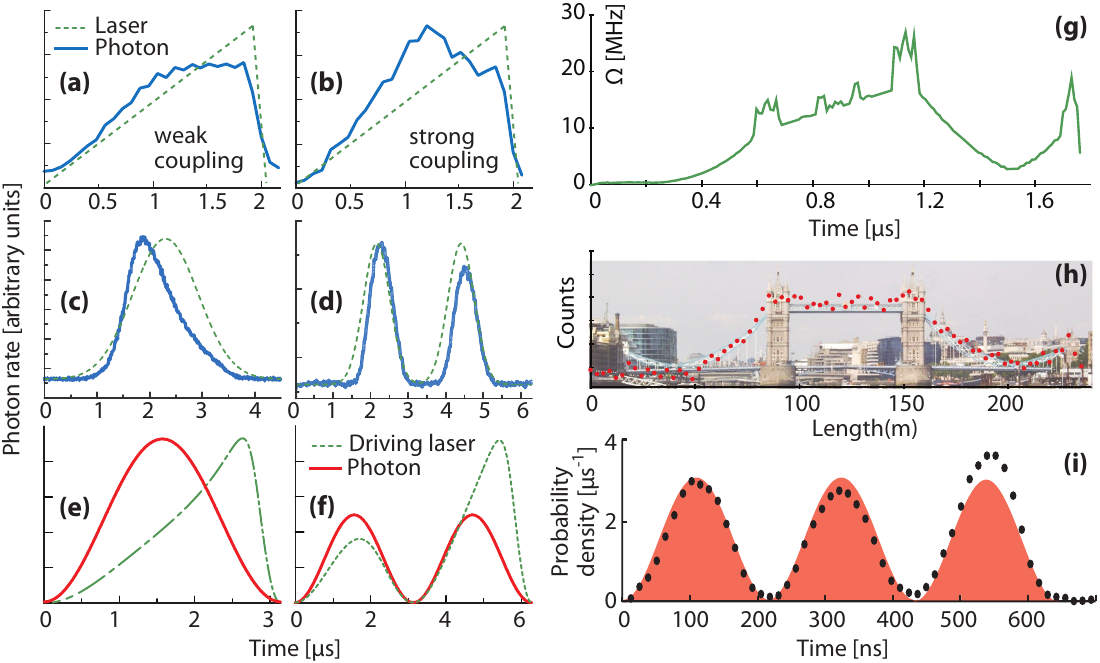}
    \caption{Photons made to measure: {\bf (a-d)} show photon shapes realised in several experiments and their driving laser pulses. The histogram of the photon-detection time has been recorded for several hundred single-photon emissions. The data shown in {\bf (a+b)} is taken from \cite{Kuhn02}, with neutral atoms falling through a high-finesse cavity acting as photon emitters. The linear increase in Rabi frequency is the same in both cases, and the difference in photon shape is caused by variations in the coupling strength to the cavity. The data shown in {\bf (c+d)} is taken from \cite{Keller04}, with a single ion trapped between the cavity mirrors. It shows that the photon shape depends strongly on the driving laser pulse
(Panels c+d adapted with permission from Nature Publishing Group: Nature, M. Keller et al. \cite{Keller04}, Copyright 2004).
{\bf (e+f)} show the Rabi frequency one needs to apply to achieve symmetric single or twin-peak photon pulses with an efficiency close to unity. This is a result from an analytical solution of the problem discussed in \cite{Vasilev10}. The latter scheme has been applied successfully for generating photons of various arbitrary shapes \cite{Nisbet11}, with examples shown in {\bf (g-i)}.\label{fig:Shape}}
\end{figure}

\keyword{Photon shaping} can be addressed by solving the master equation of the atom-photon system, which yields the time-dependent probability amplitudes, and by consequence also the wave function of the photon emitted from the cavity \cite{Kuhn02,Keller04}. Only recently, we have shown \cite{Dilley12,Nisbet11,Vasilev10} that this analysis can  be reversed, giving a unambiguous analytic expression for the time evolution of the driving field  in dependence of the desired shape of the photon. This model is valid in the strong-coupling and bad-cavity regime, and it generally allows one to fully control the coherence and population flow in any Raman process. Designing the driving pulse to obtain photonic wave packets of any possible desired shape $\psi_\mathrm{ph}(t)$ is straight forward \cite{Dilley12,Vasilev10}. Starting from the three-level atom discussed in section \ref{sec:three-level-atom}, we consider only the states $|e,0\rangle$, $|x,0\rangle$, and $|g,1\rangle$ of the $n=1$ triplet and their corresponding probability amplitudes $\mathbf{c}(t)=\left[c_\mathrm{e}(t),c_\mathrm{x}(t),c_\mathrm{g}(t)\right] ^{T}$, with the atom-cavity system initially prepared in $|e,0\rangle$. The Hamiltonian (\ref{eq:Hatcav3}) and the decay of atomic spin and cavity field at rates $\gamma$ and $\kappa$ , respectively, define the \keyword{master equation} of the system,
\begin{equation}
i \hbar \frac{d}{dt}\mathbf{c}(t)=-\frac{\hbar }{2}%
\left(
\begin{array}{ccc}
0 & \Omega (t) & 0 \\
\Omega (t) & 2i\gamma  & 2g \\
0 & 2g & 2i\kappa
\end{array}%
\right) \mathbf{c}(t).
\label{EQ:TDSE} 
\end{equation} 
The cavity-field decay at rate $\kappa$ unambiguously links the probability amplitude of $|g,1\rangle$ to the desired wave function $\psi_\mathrm{ph}(t)$ of the photon. Furthermore,  $|g,1\rangle$ only couples to $|x,0\rangle$ with the well-defined atom-cavity coupling  $g$, while the Rabi frequency $\Omega$ of the driving laser is linking $|x,0\rangle$ to $|e,0\rangle$. Hence the time evolution of the probability amplitudes and $\Omega(t)$ read
\begin{eqnarray}
c_\mathrm{g}(t)&=&\psi _\mathrm{{ph}}(t) / \sqrt{2\kappa },  \label{Cg-def}\\
c_\mathrm{x}(t)&=&-\frac{i}{g}\left[ \dot{c}_\mathrm{g}(t)+\kappa c_\mathrm{g}(t)\right] \label{Cx-eq}\\
\Omega(t)c_e(t)&=&
2  \left[ i \dot{c}_x(t)+ i \gamma  c_x(t) -g c_g(t) \right].
\end{eqnarray}
We can use the continuity of the system, taking into account the decay of atom polarisation and cavity field  at rates $\gamma$ and $\kappa$, respectively, to get to an independent expression for
\begin{equation}
|c_e(t)|^2 = 1 - |c_x(t)|^2 - |c_g(t)|^2 - \int\limits_\mathrm{0}^{t}dt\left[
2\gamma |c_x(t)|^2+2\kappa |c_g(t)|^2\right].
\end{equation}
With the Hamiltonian not comprising any detuning and assuming $\psi_\mathrm{{ph}}$ to be real, one can easily verify that the probability amplitude $c_x(t)$ is purely imaginary, while $c_e(t)$ and  $c_g(t)$ are both real.  Hence with the desired photon shape as a starting point, we have obtained analytic expressions for all probability amplitudes. These then yield the Rabi frequency
\begin{equation}
\label{eq:driving:OMT}
\Omega (t)=\frac{2  \left[ i \dot{c}_x(t)+ i \gamma  c_x(t) -g c_g(t) \right]}{\sqrt{1 - |c_x(t)|^2 - |c_g(t)|^2 - \int\limits_\mathrm{0}^{t}dt\left[
2\gamma |c_x(t)|^2+2\kappa |c_g(t)|^2\right]}},  \label{eq:Omega}
\end{equation}
which is a real function defining the driving pulse required to obtain the desired photon shape.

\begin{figure}[t]
  \centering\includegraphics[width=4.4in]{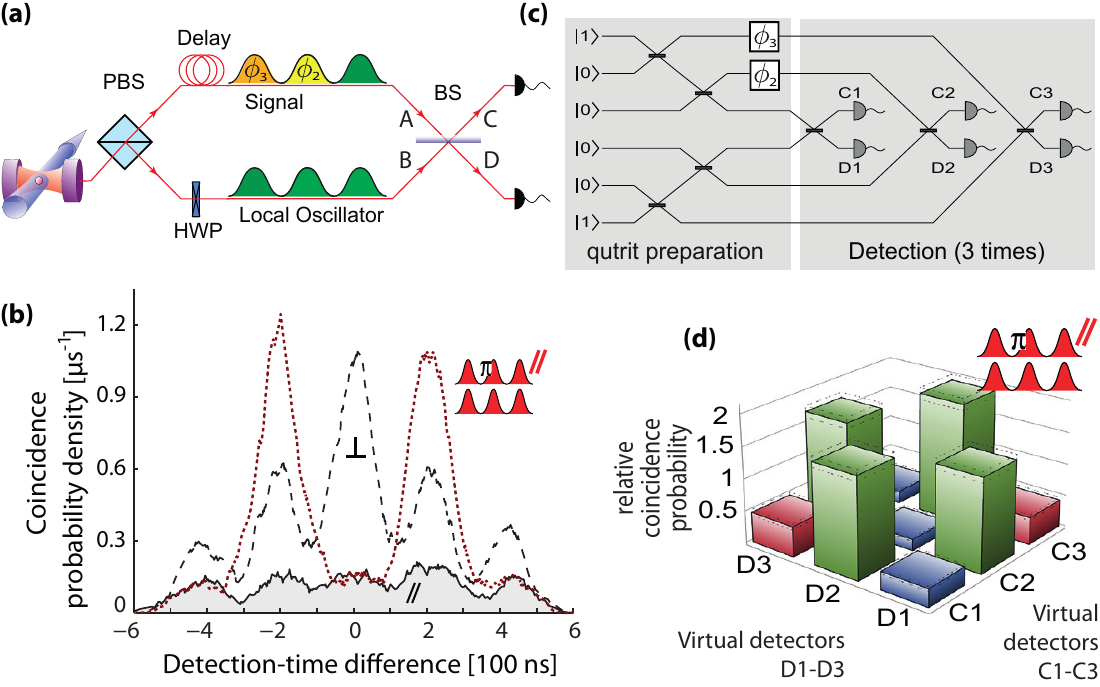}
    \caption{Qutrits, from \cite{Nisbet13}: \textbf{(a)} Pairs of triple-peak photons subsequently emitted are delayed such they arrive simultaneously at a beam splitter. Time-resolved coincidences are then registered between output ports $C$ and $D$. The signal photon carries mutual phases $\phi_\mathrm{1}$ and $\phi_\mathrm{2}$ between peaks, the local oscillator does not.
\textbf{(b)} Time-resolved homodyne signal  for photons of perpendicular (black) and parallel (blue) polarization, with the signal photon having a phase shift in the central time bin of $\phi_\mathrm{1}=\pi$ (red). The solid traces result from summing all coincidences found within a 60\,ns wide interval around each point of the trace. For some of these data points, the statistical error is shown. 
\textbf{(c)} Corresponding virtual circuit if the same experiment was done in the spatial domain. The actual physical system, consisting of one beam splitter and two detectors, would then correspond to a six-detector setup. All time-resolved photodetections in the real system can be easily associated with corresponding virtual detectors firing.   
\textbf{(d)} Relative coincidence probabilities between virtual detectors (blue: detections within the same time bin; green: detections in successive time bins; red: detections two time-bins apart).\label{fig:qudits}}
\end{figure}

Fig.\,\ref{fig:Shape} compares some of the results obtained in producing photons of arbitrary shape. For instance, the driving laser pulse shown in  Fig.\,\ref{fig:Shape}(g) has been calculated according to Eq.\,(\ref{eq:Omega}) to produce the photon shape from  Fig.\,\ref{fig:Shape}(h). From all data and calculation, it is obvious that stronger driving is required to counterbalance the depletion of the atom-cavity system towards the end of the pulse. Therefore a very asymmetric driving pulse leads to the emission of photons symmetric in time, and vice versa, as can be seen from comparing Fig.\,\ref{fig:Shape}(c) and (e).  

Amongst the large variety of  shapes that have been produced, their possible sub-division into various peaks within  separate time-bins is a distinctive feature, seen that it allows for  \keyword{time-bin encoding} of quantum information. For instance, we recently have been imprinting different mutual phases on various time bins of multi-peak photons \cite{Nisbet13}, and then successfully retrieved this phase information in a time-resolved quantum-homodyne experiment based on two-photon interference. The latter is illustrated in Fig.\,\ref{fig:qudits}. Subsequently emitted \keyword{triple-peak photons} from the atom-cavity system are sent into optical-fibre made delay lines to arrive simultaneously at a beam splitter. While the mechanism described above is used to sub-divide the photons into three peaks of equal amplitude, i.e. three well-separate time bins or temporal modes, we also impose phase changes from one time bin to the next. The latter is accomplished by phase-shifting the driving laser with an acousto-optic modulator. Therefore the signal photons emitted from the cavity are prepared in a $W$-state with arbitrary relative phases between their constituent temporal modes,
\begin{equation}
|\Psi_\mathrm{photon}\rangle = (e^{i\phi_\mathrm{1}}|1,0,0\rangle+e^{i\phi_\mathrm{2}}|0,1,0\rangle+e^{i\phi_\mathrm{3}}|0,0,1\rangle)/\sqrt{3}.
\end{equation}
We may safely assume that  $\phi_\mathrm{1}=0$ as only relative phases are of any relevance. The atom-cavity system is driven in a way that an alternating sequence of signal photons and local-oscillator photons gets emitted, with the \keyword{local-oscillator photons} not being subject to any phase shifts between its constituents, but otherwise identical to the \keyword{signal photons}. This ensures that a signal photon (including phase jumps) and a local-oscillator photon (without phase jumps) always arrive simultaneously at the beam splitter behind the delay lines whenever two successively emitted photons enter the correct delay paths. Two detectors, labeled $C$ and $D$, are monitoring the output ports of this beam splitter. Their time resolution is good enough to discriminate whether photons are detected during the first, second or third peak. The probability for photon-photon correlations across different time bins and detectors therefore reflects the phase change within the photons -- i.e. the probability for correlations between the two detectors monitoring the beam-splitter output depends strongly on the timing of the photo detections. For instance, the coincidence probability $P(C_\mathrm{i},D_\mathrm{j})$ for detector $C$ clicking in time bin $i$ and detector $D$ in time bin $j$ is then
\begin{equation}
P(C_\mathrm{i},D_\mathrm{j}) \propto \sin^2{((\phi_\mathrm{i}-\phi_\mathrm{j})/2)}.
\end{equation}
We have been exploring these phenomena  \cite{Nisbet13} with the experiment illustrated in Fig.\,\ref{fig:qudits}, using two types different signal photons. One with no mutual phase shifts, i.e. $\phi_\mathrm{1}=\phi_\mathrm{2}=\phi_\mathrm{3}=0$, and the other with $\phi_\mathrm{1}=0,\ \phi_\mathrm{2}=\pi,\ \phi_\mathrm{3}=0$. In the first case, signal and local oscillator photons are identical. By consequence, no correlations between the two detectors arise (apart from a constant background level due to detector noise). In the second case the adjacent time bins within the signal photon are $\pi$ out of phase. Therefore the probability for correlations between the two detectors increases dramatically if the detectors fire in adjacent time bins, but it stays zero for detections within the same time bin, and for detections occurring in the first and third time bin.  These new findings demonstrate nicely that atom-cavity systems give us the capability of fully controlling the temporal evolution of amplitude and phase within single deterministically generated photons. Their characterisation with time-resolved Hong-Ou-Mandel interference used for quantum homodyning the photons then reveals these phases again in the photon-photon correlations.  

The availability of time bins as an additional degree of freedom to LOQC in an essentially deterministic photon-generation scheme is a big step towards large-scale quantum computing in photonic networks \cite{Bennett08}.  Arbitrary single-qubit operations on time-bin encoded qubits seem straightforward to implement with phase-coherent optical delay lines and active optical routing to either switch between temporal and spatial modes, or to swap the two time bins. Controlling the atom-photon coupling might also allow the mapping of atomic superposition states to time-binned photons \cite{Dilley12,Ritter12}; and the long coherence time, combined with fast detectors, makes  real-time feedback possible during photon generation. 

\section{Cavity-based \keyword{quantum memories}}
Up to this point, we have been discussing cavity-based single-photon emission, atom-photon state mapping, entanglement and basic linear optical phenomena using  these photons. All these processes rely on a unitary time evolution of the atom-cavity system upon photon emission, in a process which intrinsically is fully reversible. Due to this property, it should be possible to use atoms in strong cavity coupling as universal nodes within a large quantum optical network. The latter is a very promising route towards hybrid quantum computing, which has the potential to overcome many scalability issues because it combines stationary atomic quantum bits with fast photonic links and linear optical information processing in a so-called  ``quantum internet'' \cite{Kimble08}, which is based on photon-mediated state mapping between two distant atoms placed in spatially separated optical cavities \cite{Cirac97}. Here, we basically summarize our model from \cite{Dilley12} and discuss how to expand our previous Raman scheme to capture a single photon of arbitrary temporal shape with one atom coupled to an optical cavity, using a control pulse of suitable temporal shape to ensure \keyword{impedance matching} throughout the photon arrival, which is necessary for complete \keyword{state mapping} from photon to atom.  We also note that quantum networking between two cavities has recently been experimentally demonstrated with atomic ensembles \cite{Choi08,Matsukevich06} and single atoms in strong cavity coupling \cite{Specht11,Ritter12}.

In addition to the single-photon emission and the associated quantum state mapping in emission, the newly declared goal is now to find a control pulse that achieves complete absorption of single incoming photons of arbitrary temporal shape, given by the running-wave probability amplitude $\phi_{in}(t)$, which arrives at one cavity mirror\footnote{$\phi_{in}(t)$ is the probability amplitude of the running photon, with $\int_{-\infty}^{+\infty}|\phi_{in}(t)|^2 dt =1$ and $|\phi_{in}(t)|^2dt$ the probability of the photon arriving at the mirror within $[t, t+dt]$}. This relates most obviously to mapping Fock-state encoded qubits to atomic states \cite{Cirac97,BBMEJ07}, but also extends to other possible superposition states, e.g. photonic time bin or polarisation encoded qubits \cite{Specht11,Wilk07-Science,Sun04}. 

\begin{figure}
\centering\includegraphics[width=0.95\columnwidth]{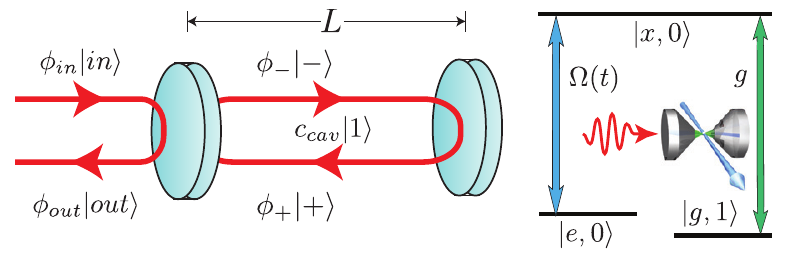}
\caption{Cavity coupling (from \cite{Dilley12}): A photon arrives from the left and either gets reflected off the cavity or couples to its internal modes. The three levels of the atom inside the cavity are labeled $|e\rangle$, $|x\rangle$ and $|g\rangle$, with photon number states $|0\rangle$ and $|1\rangle$. The couplings of the control pulse and cavity are given by $\Omega(t)$ and $g$ respectively, and the atom is initially prepared in state $|g\rangle$.}
\label{levels}
\end{figure}

Prior to investigating the effect of the atom-cavity and atom-laser coupling, we  briefly revisit the \keyword{input-output coupling} of an optical cavity in the time domain. Inside the cavity, we assume that the mode spacing is so large that only one single mode of frequency  $\omega_{cav}$ contributes, with the dimension-less probability amplitude $c_{cav}(t)$ determining the occupation of the one-photon Fock state $|1\rangle$. Furthermore, we assume that coupling to the outside field is fully controlled by the field reflection and transmission coefficients, $r$ and $\tau$, of the coupling mirror, while the other has a reflectivity of $100\,\%$. We decompose the cavity mode into submodes $|+\rangle$ and $|-\rangle$, travelling towards and away from the coupling mirror, so that the spatio-temporal representation of the cavity field reads
\begin{equation}
\phi_+(t)|+\rangle + \phi_-(t)|-\rangle,
\label{inout}
\end{equation}
where $\Delta\phi = \phi_-(t)-\phi_+(t)$ is the change of the running-wave probability amplitude at the coupling mirror.  The latter is small for mirrors of high reflectivity, such that $c_{cav}(t)\simeq \phi_+(t) \sqrt{t_r}\simeq \phi_-(t) \sqrt{t_r}$, with $t_r=2L/c$ the cavity's round-trip time.  We also decompose the field outside the cavity into incoming and outgoing spatio-temporal field modes, with running-wave probability amplitudes $\phi_{in}(t+z/c)$ and $\phi_{out}(t-z/c)$ for finding the photon in the $|in\rangle$ and $|out\rangle$ states at time $t$ and position $z$, respectively. The coupling mirror at $z=0$ acts as a beam splitter with the operator $a^\dagger_- (r a_+ + \tau a_{in}) + a^\dagger_{out} (\tau a_+ - r a_{in})$ coupling the four running-wave modes inside and outside the cavity. In matrix form, this coupling equation reads
\begin{equation}
\left(\begin{array}{c}\phi_++\Delta\phi \\ \phi_{out}\end{array}\right) 
=
\left(\begin{array}{c}\phi_-\\ \phi_{out}\end{array}\right) 
=
 \left(\begin{array}{cc}r & \tau \\ \tau & -r \end{array}\right) \left(\begin{array}{c}\phi_+\\ \phi_{in}\end{array}\right)
\label{an_cr}
\end{equation}
To relate $c_{cav}(t)$ to the running-wave probability amplitudes, we take $r\approx 1-\kappa t_r$ and $\tau=\sqrt{2\kappa t_r}$, where $\kappa$ is the field decay rate of the cavity. Furthermore, we make use of
\begin{equation}
\frac{d c_{cav}}{dt} \simeq\frac{c_{cav}(t+t_r)-c_{cav}(t)}{t_r}= \frac{(\phi_--\phi_+)\sqrt{t_r}}{t_r} = \frac{\Delta\phi}{\sqrt{t_r}}.
\end{equation}
With these relations, the first line of Eq.\,(\ref{an_cr}) can be written as $c_{cav}/t_r + \dot{c}_{cav} = (1-\kappa t_r) c_{cav}/t_r +\sqrt{2\kappa}\phi_{in}$. Therefore Eq.\,(\ref{an_cr}) takes the form of a differential equation
\begin{equation}
\left(
\begin{array}{c}\dot{c}_{cav}\\ \phi_{out}\end{array}\right) = \left(\begin{array}{cc}-\kappa & \sqrt{2\kappa}\\ \sqrt{2 \kappa} & -r \end{array}\right) \left(\begin{array}{c}c_{cav}\\ \phi_{in}\end{array}\right).
\label{inputoutput}
\end{equation}
This describes the coupling of a resonant photon into and out of the cavity mode. The reader might note that the result obtained from our simplified input-output model is fully equivalent to the conclusions drawn from the more sophisticated standard approach that involves a decomposition of the continuum into a large number of frequency modes \cite{Walls-Milburn}. No such decomposition is applied here as we study the problem uniquely in the time domain. 

Next, we examine the coupling of a single atom to the cavity, as discussed in the preceding sections. We again consider a three level $\Lambda$-type atom with two electronically stable ground states $|e\rangle$ and $|g\rangle$, coupled by either the cavity field mode or the control laser field to one-and-the-same electronically excited state $|x\rangle$. For the one-photon multiplet of the generalised Jaynes-Cummings ladder, the cavity-mediated coupling between $|g,1\rangle$ and $|x,0\rangle$ is given by the atom-cavity coupling strength $g$, while the control laser  couples $|e,0\rangle$ with $|x,0\rangle$ with Rabi frequency $\Omega(t)$. The probability amplitudes of these particular three product states read $c_e(t)$, $c_g(t)$, and $c_x(t)$, respectively, with their time evolution  given by 
\begin{equation}
\setlength\arraycolsep{1.4pt}
\left(
\begin{array}{c}\dot{c}_e \\ \dot{c}_x \\ \dot{c}_g \\ \phi_{out}\end{array}\right)=
 \left(\begin{array}{cccc}
0 & -i \Omega(t)^*/2 & 0 & 0\\
-i \Omega(t)/2 & -\gamma & -i g & 0\\ 
0 & -i g^* & -\kappa & \sqrt{2\kappa} \\ 
0&0 & \sqrt{2\kappa} & -r \end{array}
\right) \left(\begin{array}{c}c_e\\ c_x\\ c_g\\ \phi_{in}\end{array}\right),
\label{TDSE}
\end{equation}
which is formally equivalent to the Schr\"{o}dinger equation (\ref{EQ:TDSE}) which we've been considering for single-photon shaping, but now includes the input-output relation from Eq.\,(\ref{inputoutput}). This new \keyword{master equation} is modelling the atom coupled  to the cavity as an open quantum system, driven by the incoming photon, with its probability amplitude  $\phi_{in}(t)$ to be taken at $z=0$, and possibly coupling or directly reflecting light into the outgoing field with amplitude $\phi_{out}(t)$. We also note that $c_g(t) \equiv c_{cav}(t)$  because the state $|g,1\rangle$ is the only atom-field product state in which there is one photon in the cavity.  On resonance, $g$ and $\phi_{in}(t)$ are both real, and by consequence $c_g(t)$ and $c_e(t)$ are real while  $c_x(t)$ is purely imaginary. Because we are considering only one photon, the probability of occupying $|g,0\rangle$ is given by the overall probability of having a photon outside the cavity, either in state $|in\rangle$  or in $|out\rangle$.  These states couple only via the cavity mirror to $|g,1\rangle$. 

The realisation of a cavity-based single-atom quantum memory is based on the complete absorption of the incoming photon, which is described by its time-dependent probability amplitude $\phi_{in}(t)$. This calls for perfect impedance matching, i.e. no reflection and $\phi_{out}(t)=0$  at all times. This condition yields
\begin{eqnarray}
c_g(t) &=&\phi_{in}(t) / \sqrt{2\kappa} \label{cgg}\\
c_x(t) &=&i  \left[ \dot{c}_g(t)-\kappa  c_g(t) \right]  /g^* = i  \left[ \dot{\phi}_{in}(t)-\kappa  \phi_{in}(t) \right]  /g^*\sqrt{2\kappa}\\ 
\Omega(t)c_e(t)&=&
2    \left[ i \dot{c}_x(t)+ i \gamma  c_x(t) -g c_g(t) \right]. 
\label{OCE}\label{cxx}
\end{eqnarray}
With the photon initially completely in the incoming state $|in\rangle$, i.e. $\int|\phi_{in}(t)|^2dt = 1$, and the atom-cavity system in state $|g,0\rangle$, the continuity balance yields
\begin{equation}
|c_e(t)|^2=|c_0|^2-|c_g(t)|^2-|c_x(t)|^2+\int_{-\infty}^t[|\phi_{in}(t')|^2 - 2\gamma |c_x(t')|^2 ]dt',
\label{REE}
\end{equation}
including an offset term $|c_0|^2$ to account for a small non-zero initial occupation of $|e,0\rangle$. The relevance of this term becomes obvious in the following. From Eqs. (\ref{OCE}, \ref{REE}), we  obtain the  Rabi frequency of the driving pulse,
\begin{equation}
\Omega(t)=\frac{2 \left[ i \dot{c}_x(t)+ i \gamma  c_x(t) -g c_g(t) \right]}
{\sqrt{|c_0|^2-|c_g(t)|^2-|c_x(t)|^2+\int_{-\infty}^t[|\phi_{in}(t')|^2 - 2\gamma |c_x(t')|^2 ]dt'}},
\label{OMT}
\end{equation}
necessary for full impedance matching over all times. In turn this assures complete absorption of the incoming photon by the atom-cavity system. We emphasize here the close similarity of this novel expression with the analytic form of the driving pulse needed for the emission of shaped photons, Eq.\,(\ref{eq:driving:OMT}). This is not a coincidence. The photon absorption discussed here is nothing else than a time-reversal of the photon emission process. If we disregard any losses and also assume the final occupation of $|e,0\rangle$ after a photon emission equals the initial occupation of that state prior to a photon absorption, then the required Rabi frequency $\Omega(t)$ for absorption is the exact mirror image in time of the Rabi frequency used for photon generation. 
 
Let us now consider physically realistic photons restricted to a finite support of well-defined start and end times, $t_{start}$ and $t_{stop}$, starting smoothly with $\phi_{in}(t_{start})=\frac{d}{dt}{\phi}_{in}(t_{start})=0$, but of non-zero second derivative. Therefore Eq. (\ref{OCE}) yields $\Omega(t_{start})c_e(t_{start})\neq 0$. This necessitates a small initial population in state $|e,0\rangle$ because otherwise perfect impedance matching with $|c_0|^2= 0$ is only possible with photons of infinite duration. 

\begin{figure}
\centering\includegraphics[width=0.95\columnwidth]{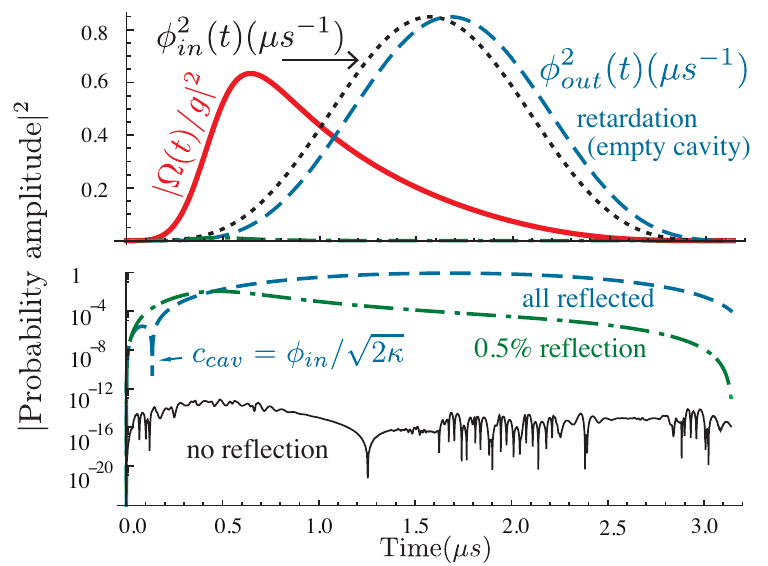}
\caption{Impedance matching (from \cite{Dilley12}): Incoming sin$^{2}$ photon (dotted). Case (a): Empty cavity, all reflected (dashed); Case (b): System prepared in $|g,0\rangle$, small reflection (dash-dotted); Case (c): Small initial population in $|e,0\rangle$, reflection suppressed (thin solid). The control pulse (thick solid) is derived to match case (c).}
\label{sin2}
\end{figure}

To illustrate the power of the procedure and the implications of the constraints to the initial population, we  apply the scheme to a typical photon shapes that one may obtain from atom-cavity systems. We consider a cavity with parameters similar to one of our own experimental implementations, with $(g,\kappa,\gamma)=2\pi\times (15, 3, 3)\,$MHz and a resonator length of $L=100\,\mu$m.
As an example, we assume that a symmetric photon with $\phi_{in}(t)\propto \sin^2(\pi t/\tau_{photon})$ arrives at the cavity. For a photon duration of $\tau_{photon}=3.14\,\mu$s, fig.\,\ref{sin2} shows $\phi_{in}(t)$,  $\Omega(t)$  and the probability amplitude of the reflected photon, $\phi_{out}(t)$, as a function of time. The latter is obtained from a numerical solution of eqn.\,(\ref{TDSE}) for the three cases of (a) an empty cavity, (b) an atom coupled to the cavity initially prepared in $|g,0\rangle$, with $|c_0|=0$, and (c) a small fraction of the atomic population initially in state $|e,0\rangle$, with $|c_0|^2=0.5\%$. In all three cases, the Rabi frequency $\Omega(t)$ of the control pulse is identical. It has been calculated analytically assuming a small value of $|c_0|^2=0.5\%$ (this choice is arbitrary and only limited by practical considerations, as will be discussed later).  From these simulations, it is obvious the photon gets fully reflected if no atom is present (case a), albeit with a slight retardation due to the finite cavity build-up time. Because the direct reflection of the coupling mirror is in phase with the incoming photon and the light from the cavity coupled through that mirror is out-of phase by $\pi$, the phase of the reflected photon flips around as soon as $c_{cav}(t)=\phi_{in}(t)/\sqrt{2\kappa}$. This shows up in the logarithmic plot as a sharp kink in $\phi_{out}(t)$ around $t=0.13\,\mu$s.

The situation changes dramatically if there is an atom coupled to the cavity mode. For instance, with the initial population matching the starting conditions used to derive $\Omega(t)$, i.e. case (c) with $\rho_0=0.5\%$, no photon is reflected. The amplitude of $|\phi_{out}(t)|^2$ remains below $10^{-12}$, which corresponds to zero within the numerical precision. However, for the more realistic case (b) of the atom-cavity system well prepared in $|g,0\rangle$,  the same control pulse is not as efficient, and the photon is reflected off the cavity with an overall probability of $0.5\%$. This matches the ``defect'' in the initial state preparation, and can be explained by the finite cavity build-up time leading to an impedance mismatch in the onset of the pulse.  

We emphasise that this seemingly small deficiency in the photon absorption might become significant with photons of much shorter duration. For instance, in the extreme case of a photon duration $\tau_{photon}<\kappa^{-1}$, building up the field in the cavity to counterbalance the direct reflection by means of destructive interference is achieved most rapidly without any atom. Any atom in the cavity will act as a sink, removing intra-cavity photons. With an atom present, a possible alternative is to start off with a very strong initial Rabi frequency of the control pulse. This will project the atom-cavity system initially into a dark state, so that the atom does not deplete the cavity mode. Nonetheless, the initial reflection losses would still be as high as for an empty cavity.  

\begin{figure}[h]
\centering\includegraphics[width=0.95\columnwidth]{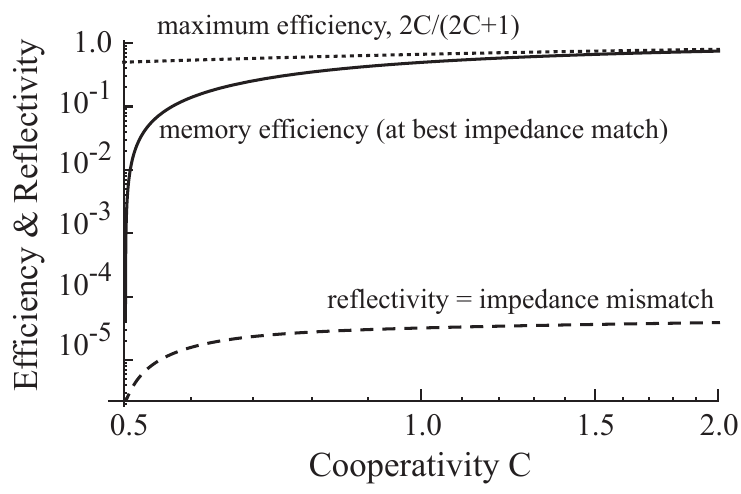}
\caption{Storage efficiency (mapping the photon to $|e,0\rangle$, solid line) and impedance mismatch (back-reflecting the photon, dashed line) as a function of the cooperativity, $C=g^2/(2\kappa\gamma)$ (from \cite{Dilley12}). The dotted line shows the maximum possible efficiency. Both efficiency and impedance mismatch have been numerically calculated for $\kappa=\gamma=2\pi 3\,$MHz using symmetric $\sin^2$ pulses of $3.14\,\mu$s duration, with control pulses modelled according to Eqs.\.(\ref{cxx}-\ref{OMT}) to achieve optimum impedance match.}
\label{lossy}
\end{figure}

To illustrate the interplay of impedance matching and \keyword{memory efficiency}, Fig.\,\ref{lossy} shows the reflection probability and the memory efficiency (excitation transfer to $|e,0\rangle$) as a function of  the \keyword{cooperativity}, $C=g^2/(2\kappa\gamma)$. Obviously, the impedance matching condition is always met, but the efficiency varies. For $C>1$, it asymptotically reaches the predicted optimum \cite{Gorshkov07} of $2C/(2C+1)$, but it drops to zero at $C=1/2$ (i.e. for $g=\kappa=\gamma$). In this particular case, the spontaneous emission loss via the atom equals the transmission of the coupling mirror. Hence the coupled atom-cavity system behaves like a balanced Fabry-Perot cavity, with one real mirror being the input coupler, and the spontaneously emitting atom acting as output coupler. Therefore the photon goes into the cavity, but is spontaneously emitted by the atom and gets not mapped to $|e,0\rangle$. This limiting case furthermore implies that impedance matching is not possible for $C<1/2$, as the spontaneous emission via the atom would then outweigh the transmission of the coupling mirror. The application of our formalism therefore fails in this weak coupling regime  (actually, the evaluation of Eq.\,(\ref{REE}) would then yield values of  $|c_e|^2<0$, which is not possible).

The photon reabsorption scheme discussed here, together with the earlier introduced method for generating tailored photons \cite{Nisbet11,Vasilev10,Nisbet13}, constitute the key to analytically calculating  the optimal driving pulses needed to produce and absorb arbitrarily shaped single-photons (of finite support) with three level $\Lambda$-type atoms in optical cavities. This is a sine qua non condition for the successful implementation of a quantum network.  It is expected that this simple analytical method will have significant relevance for those striving to achieve atom-photon state transfer in cavity-QED experiments, where low losses and high fidelities are of paramount importance.

\section{Future directions}
We have discussed a large variety of ways for producing single photons from simple quantum systems. The majority of these photon-production methods lead to on-demand emission of narrowband and indistinguishable photons into a well defined mode of the radiation field, with efficiencies that can be very close to unity. Therefore these photons are ideal for all-optical quantum computation schemes, as proposed by Knill, Laflamme, and Milburn  \cite{Knill01}. These sources are expected to play a significant role in the implementation of quantum networking \cite{Cirac97} and quantum communication schemes \cite{Briegel98}.

The atom- and ion-based sources have already shown to be capable of entangling and mapping quantum states between atoms and photons \cite{Wilk07-Science,Weber09}. Processes like entanglement swapping and teleportation between distant atoms or ions, that have first been studied without the aid of cavities \cite{Blinov04,Volz06,Sun04,Beugnon06,Maunz07} are beginning to benefit enormously from the introduction of cavity-based techniques \cite{Specht11,Ritter12,Nolleke13}, as their success probability scales with the square of the efficiency of the photon generation process. The high efficiency of cavity-based photon sources also opens up new avenues towards a highly scalable quantum network, which is essential for providing cluster states in one-way quantum computing \cite{Raussendorf01} and for the quantum simulation of complex solid-state systems \cite{Buechler05}.

\begin{acknowledgement}
Hereby I express my gratitude to all my colleagues and co-workers in my present and past research groups at the University of Oxford and the MPQ in Garching. It has been their engagement and enthusiasm, teamed up with the support from the European Union, the DFG and the EPSRC, which lead to the discovery of the phenomena and the development of the techniques discussed in this chapter. 
\end{acknowledgement}

\bibliographystyle{SpringerPhysMWM} 

\printindex
\end{document}